\documentclass[preprint2]{aastex}

\slugcomment{to appear in AJ.}

\shorttitle{Globular Clusters in NGC 3115 DW1}
\shortauthors{Puzia et al.}

\begin{document}
\title{Globular Clusters in the dE,N galaxy NGC 3115 DW1: New Insights from
Spectroscopy and HST Photometry}

\author{Thomas H. Puzia} \affil{Sternwarte der Universit\"at Bonn, Auf
  dem H\"ugel 71, 53121 Bonn, Germany}
\email{tpuzia@astro.uni-bonn.de}
\author{Markus Kissler-Patig} \affil{European
  Southern Observatory, Karl-Schwarzschild-Strasse 2, 85748 Garching
  bei M\"unchen, Germany} 
\email{mkissler@eso.org}
\author{Jean P. Brodie \& Linda L. Schroder}\affil{UCO/Lick Observatory,
  University of California at Santa Cruz, Santa Cruz, CA 95064}
\email{brodie@ucolick.org, linda@ucolick.org}

\begin{abstract}
  The properties of globular clusters in dwarf galaxies are key to
  understanding the formation of globular cluster systems, and in
  particular in verifying scenarios in which globular cluster systems
  of larger galaxies formed (at least partly) from the accretion of
  dwarf galaxies. Here, we revisit the globular cluster system of the
  dE,N galaxy NGC 3115 DW1 -- a companion of the nearby S0 galaxy NGC
  3115 -- adding Keck/LRIS spectroscopy and HST/WFPC2 imaging to
  previous ground-based photometry. Spectra for seven globular
  clusters reveal normal abundance ratios with respect to the Milky
  Way and M31 clusters, as well as a relatively high mean metallicity
  ([Fe/H]$\approx -1.0\pm0.1$ dex). Crude kinematics indicate a high
  velocity dispersion within 10 kpc which could either be caused by
  dark matter dominated outer regions, or by the stripping of outer
  globular clusters by the nearby giant galaxy NGC 3115. The total
  galaxy mass out to 3 and 10 kpc lies between $1\cdot10^{10}$ and
  $1\cdot10^{11}$M$_\odot$ and $2\cdot10^{10}$ and $4\cdot10^{11}$
  M$_\odot$, respectively, depending on the mass estimator used and
  the assumptions on cluster orbits and systemic velocity. The HST
  imaging allows measurement of sizes for two clusters, returning core
  radii around 2.0 pc, similar to the sizes observed in other
  galaxies. Spectroscopy allows an estimate of the degree of
  contamination by foreground stars or background galaxies for the
  previous ground-based photometry, but does not require a revision of
  previous results: NGC 3115 DW1 hosts around $N_{\rm GC}=60\pm20$
  clusters which corresponds to a specific frequency of $S_{\rm
  N}=4.9\pm1.9$, on the high end for massive dEs. Given its absolute
  magnitude ($M_V=-17.7$ mag) and the properties of its cluster
  system, NGC 3115 DW1 appears to be a transition between a luminous
  dE and low-luminosity E galaxy.

\end{abstract}


\section{Introduction}

The study of globular cluster systems of dwarf galaxies complements
the numerous studies of such systems in larger elliptical and spiral
galaxies. Few globular cluster systems (GCSs) around dwarf galaxies
beyond the Local group have been studied to date with respect to their
cluster system (see Ashman \& Zepf 1998). This is mostly due to the
low numbers of globular clusters present in such galaxies. However,
their properties are relevant for a number of globular cluster system
formation scenarios. Dwarf galaxies are expected to provide insight
into how the smallest galaxies build up a system of globular clusters.
Further, their properties must be known in order to verify scenarios
in which larger globular cluster systems are predicted to build up by
the accretion of proto-galactic fragments or dwarf galaxies
(\citeauthor{kisslerpatig98} 1998a, \citeauthor{cote98} 1998,
\citeauthor{hilker99} 1999). These scenarios relate to the older idea
that galaxy halos might have formed through the assembly of such small
stellar systems \citep[e.g.][]{searlezinn78}.

Photometric studies of several globular cluster systems in dwarf
galaxies were carried out by \citet{durrell96b} and \citet{miller98}.
\citet{durrell96b} studied the systems of 11 dwarf galaxies in the
Virgo cluster. All were found to host globular cluster candidates and
have specific frequencies ranging from 3 to 8, similar to Local Group
dwarfs and giant elliptical galaxies. \citet{miller98} studied 24
dwarf ellipticals in the Virgo and Fornax clusters as well as in the
Leo group. They found that dE,N galaxies had higher specific
frequencies than dE galaxies, with values around $S_{\rm
  N}=6.5\pm1.2$, increasing with increasing $M_V$ (decreasing
luminosity). Not much is known yet about the metallicities of globular
clusters in dwarf galaxies. \citet{minniti96} constructed a
metallicity distribution for all Local Group dwarf galaxies and
noticed that the distribution was peaked around [Fe/H]$\approx -1.7$
dex with no clusters more metal-rich than [Fe/H]$=-1.0$ dex.
\citet{durrell96b} derived metallicities from Washington colors for
two of their Virgo dwarf ellipticals and obtained a mean metallicity
of [Fe/H]$=-1.45\pm0.2$ dex.

\citet{durrell96} studied the GCS of the dE NGC 3115 DW1 in more
detail and found it to be relatively rich (see below for a more
detailed description of their results). This motivated us to carry out
spectroscopy for some of the globular cluster candidates in this
galaxy to get a more detailed picture of their chemical and
kinematical properties. Further, HST/WFPC2 data were available from
the archive, allowing us to study the sizes of some of the clusters.

NGC 3115 DW1 is a dE1,N galaxy in the vicinity of the giant S0 galaxy
NGC 3115. It is located RA: 10h 05m 41.6s; Dec: $-07^{\rm o}$
58\arcmin\ 53.5\arcsec\ ($l=248.12^{\rm o}$; $b=36.69^{\rm o}$). We
will assume a distance of $11^{+5.0}_{-2.3}$ Mpc throughout this paper
following \citet{durrell96}. Additional properties will be given in
the text where they are relevant. In Section 2 we describe the new
data. In Section 3 we analyze the spectroscopic data giving first a
brief kinematical study of the globular cluster system before
discussing the abundances and the overall metallicity of the system.
In Section 4 we revisit the previous photometry (number of clusters,
colors, specific frequency), compare the photometric metallicities
with the spectroscopic ones and add the HST/WFPC2 imaging to derive
sizes for two clusters. In Section 5 we discuss whether NGC 3115 DW1
could have suffered stripping by its giant companion. We summarize our
results in Section 6.

\section{Data}

\subsection{Spectroscopic Observation}
\label{ln:specobs}

46 spectra of candidate globular clusters in NGC 3115 DW1 were
obtained with the LRIS-A spectrograph \citep{oke95} on the Keck-II
telescope during the nights of 1996 December 18th and 19th. The
observations were performed in multi-slit mode using two slit masks
containing 24 (mask A) and 22 slits (mask B), respectively. Mask A was
exposed for 4800 seconds while mask B was exposed 6600 seconds. The
multi-object-spectra images were taken with a Tektronix
2048$\times$2048 pix$^{2}$ chip. The seeing was around 0.8\arcsec\ 
during both nights. For all observations the first order of the
600/5000 grism (i.e.~600 lines mm$^{-1}$) was used with slitlets of
1\arcsec\ which provided a dispersion of 1.3 \AA\ pix$^{-1}$ (at 5000
\AA) and a spectral resolution of 4.1 \AA\ FWHM$^{-1}$.

All frames were binned (1$\times2$) perpendicular to the dispersion
axis during the read-out. For each night the images were reduced
individually and eventually combined. A master bias was created from 5
zero-second images, taken at the end of each night. A master flat was
obtained from 5 twilight-flat images obtained each night. Each object
image was bias subtracted and flat-fielded in a standard way. With the
IRAF\footnote{IRAF is distributed by the National Optical Astronomy
  Observatories, which are operated by the Association of Universities
  for Research in Astronomy, Inc., under cooperative agreement with
  the National Science Foundation.} package {\it apall} we traced,
extracted and sky subtracted all object spectra. The sky was
``optimally'' subtracted, i.e.~modeled with variance weighting
perpendicular to the object spectra before being subtracted
\citep{horne86}. HgKrNe-calibration-lamp spectra were obtained for
wavelength calibration. The calibration spectra were traced,
extracted, and sky subtracted exactly in the same way as the object
spectra. The wavelength calibration was verified on sky spectra
included in each slit spectrum. The overall standard deviation of the
wavelength calibration was determined to be $\sigma_{\rm
  cal}\leq0.12$\AA\ for both nights.

\begin{figure}
\epsscale{1.0}
\plotone{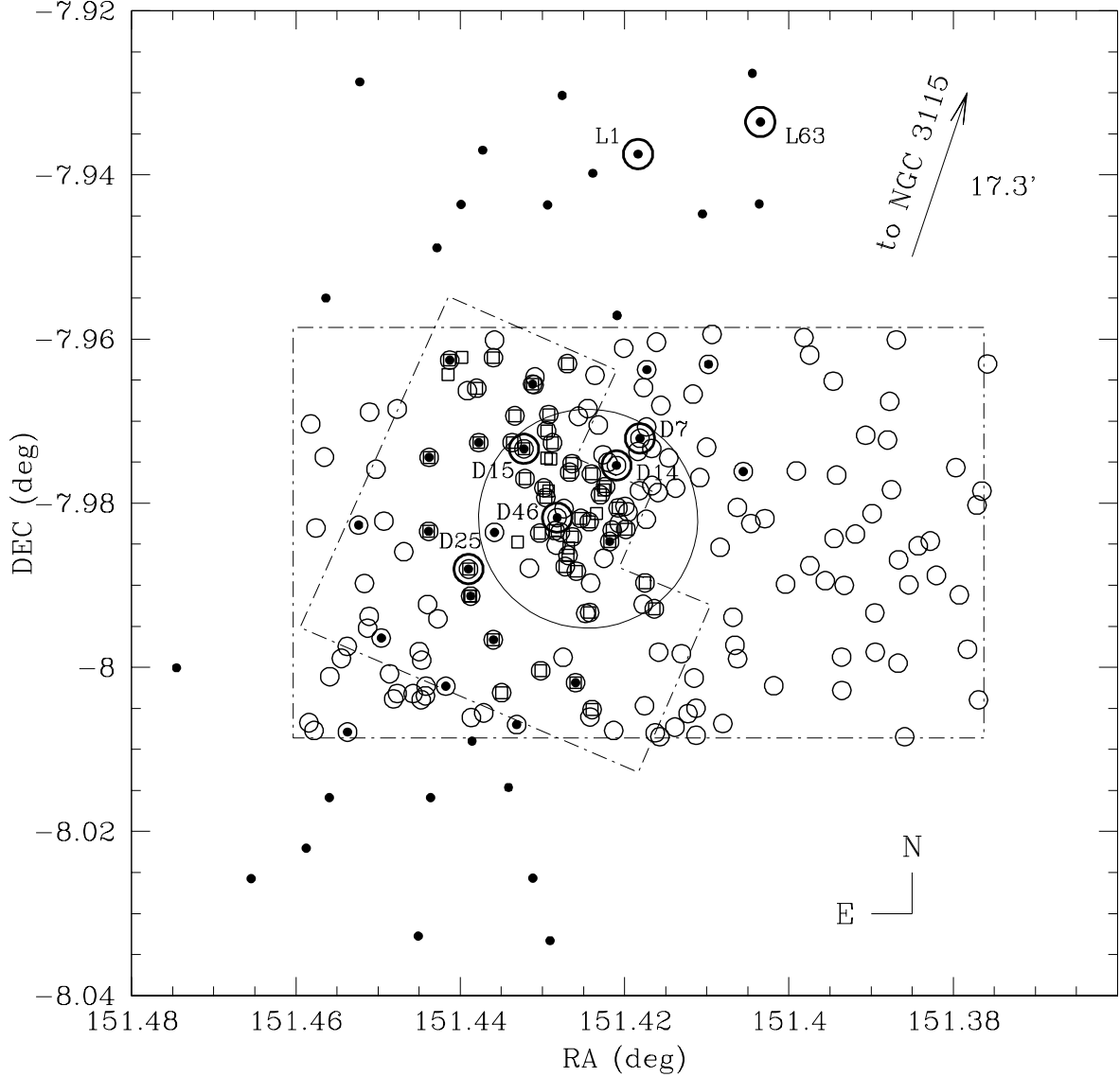}
\caption{The alignment of the field-of-views (FOV) of photometry data
and all spectroscopically analyzed objects around NGC 3115 DW1 is
shown. Dots mark spectroscopically studied objects (this paper). Open
squares are objects which were detected in a HST image (this paper)
while open circles are objects for which photometry was obtained by
\citet{durrell96}. Bold circles mark spectroscopically confirmed
globular clusters. The dot-dashed rectangle shows the FOV of the
\citet*{durrell96} photometry while the dot-dashed L-shaped FOV
belongs to the HST photometry. A large solid circle indicates the
location of $r=48$\arcsec\ radius at which the object overdensity
drops to background value \citep{durrell96}. The distance from the
center of NGC 3115 DW1 to NGC 3115 is 17.3\arcmin\ (55 kpc). The
direction is indicated by the arrow in the upper right corner. North
is up, east is left.}
\label{ps:fov}
\end{figure}

The field-of-view of LRIS-A is 6\arcmin\ $\times$ 8\arcmin . We placed
the multi-slit masks on the central/north and south-eastern part of
NGC 3115 DW1 as can be seen in Fig.~\ref{ps:fov}. The slit masks were
aligned such that the slits were pointing towards the center of the
giant S0 galaxy NGC 3115. Due to the restrictions in making the masks
(long enough slits, wavelength coverage, etc.) about 50\% of the
targets could be selected from the previous photometry of
\citet{durrell96} (see Fig.~\ref{ps:fov}). The other 50\% was
`blindly' selected from an LRIS acquisition image. Only 6 of the
former group lay within 48\arcsec\ of the galaxy center, i.e.~within
the region where \citet{durrell96} detected a clear overabundance of
objects. We therefore expected $\sim$ 6+ objects in our total sample
to be true globular clusters.

\subsection{Additional Photometry}

In order to measure sizes of globular cluster candidates that were
confirmed by our spectroscopy, additional data for the globular
cluster system were obtained from the HST archive\footnote{ Based on
  observations made with the NASA/ESA Hubble Space Telescope, obtained
  from the data archive at the Space Telescope Science Institute.
  STScI is operated by the Association of Universities for Research in
  Astronomy, Inc. under NASA contract NAS 5-26555.}. Short exposures
of 160 seconds in F555W and 320 seconds in F814W filter were taken
from program GO:5999 (PI: A.Phillips). A simple reduction procedure
was applied to all HST images: The images were biased, flat-fielded
and calibrated as described in \citet{holtzman95}, including
corrections accounting for CTE and field distortions. The brightest
clusters were matched on both the F555W and F814W images and their
sizes were studied. In addition, we re-analyzed the ground-based B,V
data\footnote{The data were kindly provided in electronic form by
  Patrick Durrell.} from \citet{durrell96}. The field-of-views of both
photometric datasets are over-plotted in Fig.~\ref{ps:fov}.

\section{Analysis of Spectroscopic Data}

In this section we extract heliocentric radial velocities $v_{\rm
  rad}$ of globular clusters and perform a mass estimate of the host
galaxy, NGC 3115 DW1. Subsequently, we measure abundances using the
Lick/IDS passband definitions and infer a mean metallicity of the GCS.

\subsection{Kinematics}

We derived radial velocities by two different methods. Velocities were
derived by cross-correlation with high S/N template spectra of two
bright GCs in M31: 225-280 and 158-213 \citep[for nomenclature
see][]{huchra82}. The cross-correlation was carried out with the IRAF
task {\it fxcor}. All measurements are summarized in Table
\ref{tab:rvs}. The ``internal'' errors which are given by the
cross-correlation code lie about $\sigma\sim 80-100$ km s$^{-1}$ for
well defined high-S/N spectra and increase rapidly as the object
spectrum becomes less defined.

For 7 high-S/N spectra we also estimated the radial velocity by
measuring redshifts of individual absorption lines (see footnote in
Table \ref{tab:rvs}). The mean radial velocity from cross-correlation
served as a first guess to the mean wavelength-shift determination. We
used the IRAF package {\it rvidlines} which employs a {\it center1d}
code to match the center of each individual absorption feature (see
manual of {\it rvidlines}). We obtained an ``internal'' mean error
from the averaging process which is $\sim$ 100 km s$^{-1}$. The
results and their errors are included in Table \ref{tab:rvs}.

\subsubsection{Selection of Globular Clusters}

After obtaining the radial velocities from cross-correlation and in
some cases from wavelength shifts of individual absorption lines, we
combined all available measurements for further selection of globular
clusters by radial velocity. Velocities derived from cross-correlation
were given twice the weight of velocities from absorption lines. The
average radial velocity for each object can be found in the last but
one column of Table \ref{tab:rvs}.

\begin{figure}
\epsscale{1.0}
\plotone{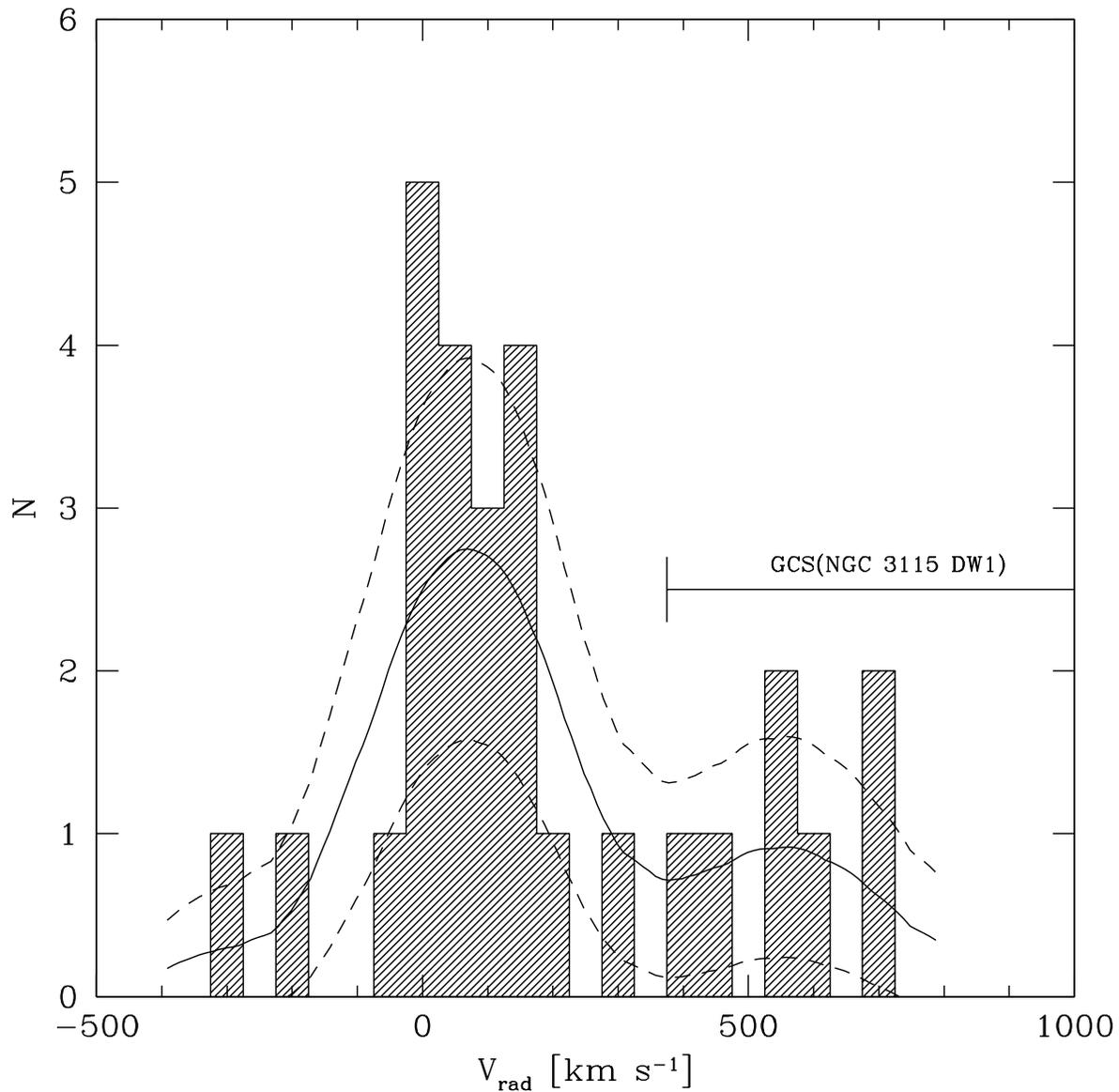}
\caption{Histogram of radial velocities. Only objects with a radial
velocity of $-500\leq v_{\rm rad}\leq1000$ km s$^{-1}$ have been
plotted. The radial-velocity range in which globular clusters are
expected is indicated (object D41 with $v_{\rm rad}=308\pm45$ km
s$^{-1}$ deviates by more than four times the velocity dispersion of
the globular cluster system from the mean system velocity of NGC 3115
DW1; it was therefore dropped). The solid line is an
Epanechnikov-kernel density estimation with a kernel width of 100 km
s$^{-1}$ \citep[for details see][]{silverman86}. Its 1-$\sigma$
uncertainty is mark by the dashed lines.}
\label{ps:rvhisto}
\end{figure}

Figure \ref{ps:rvhisto} shows a radial-velocity histogram of objects
with $-500\leq v_{\rm rad}\leq1000$ km s$^{-1}$. The figure shows a
hint of bimodality due to the extended tail of velocities $>$ 500 km
s$^{-1}$. Based on the measured radial velocity of NGC 3115 DW1 of
$\sim 700$ km s$^{-1}$ ($715\pm62$ km s$^{-1}$ \citeauthor{RC3} 1991,
$716\pm19$ km s$^{-1}$ \citeauthor{peterson93} 1993, and $698\pm42$ km
s$^{-1}$ \citeauthor{capaccioli93} 1993) and an adopted velocity
dispersion of the GCS in NGC 3115 DW1 of $\sim100$ km s$^{-1}$ (see
below), globular clusters should have radial velocities in the range
$400-1000$ km s$^{-1}$. Seven candidates lie within this range. This
cut enables a reliable differentiation between foreground objects of
low radial velocity and globular clusters in NGC 3115 DW1.

We used the KMM code \citep{ashman94} to obtain the statistical
significance of bimodality. The KMM code fits two Gaussians to the
data using Maximum-Likelihood techniques. The data can be fit with two
Gaussians of identical dispersion or two Gaussians of independent
dispersion. We applied both techniques. The mean radial velocities of
the two Gaussians are $48\pm26$ km s$^{-1}$ and $567\pm46$ km s$^{-1}$
for identical-dispersion and $52\pm28$ km s$^{-1}$ and $575\pm39$ km
s$^{-1}$ for independent-dispersion fitting. The error is the error of
the mean calculated from the variance of each distribution. KMM
estimates also the fraction of data points which are part of each
sub-distribution. Of 28 objects in the histogram (one additional
object, D26, is not included due to $v_{\rm rad}<-500$ km s$^{-1}$) in
Figure \ref{ps:rvhisto}, KMM assigns 21 objects to the sub-population
with the lower mean radial velocity and 7 to the sub-population with
the higher mean radial velocity which is in good agreement with the
expectations (see \ref{ln:specobs}). The confidence level for
bimodality is $>99$\%.

The (weighted) mean radial velocity of all 7 detected globular
clusters is $v_{\rm rad}=572\pm 30$ km s$^{-1}$. This value deviates
from the measured radial velocity of NGC 3115 DW by $\sim1.6\sigma$
which is a hint that we are biased towards lower velocities by both
the small sample and the choice of slit-mask position on the
sky. Since we detected globular clusters predominantly in the central
and northern field of NGC 3115 DW1 our measurements could be
influenced by a systematic rotation of the GCS (see also
Sect.~\ref{ln:rotdisp}).

All 22 objects with low radial velocities have a weighted mean
velocity of $v_{\rm rad}=50\pm19$ km s$^{-1}$. Assuming a simple
stellar rotation model for the Milky Way we expect the foreground
stars in the direction of NGC 3115 DW1 ($l=248.12^{\rm o}$ and
$b=36.69^{\rm o}$) to have $v_{\rm rad}=220\cdot\sin2l\cos^2b\; =98$
km s$^{-1}$ \citep{vandekamp67} which is in rough agreement with our
measurement.

In summary, within our data of 46 spectros\-copically-analyzed objects
we found 7 globular clusters 22 foreground stars, and 15 background
galaxies (9 of them are significantly clumped about $v_{\rm
  rad}\approx24000$ km s$^{-1}$ or $z\approx0.08$; the velocity
dispersion of this potential galaxy cluster is $\sigma=1300$ km
s$^{-1}$). 2 objects (D21,L6) could not be identified reliably and
were therefore dropped. The radial-velocity data for all objects are
summarized in Table \ref{tab:rvs}.

\subsubsection{Mass Estimate of NGC 3115 DW1}
\label{ln:massestimate}

Our data sample of 7 globular clusters is only sufficient for a first
rough mass estimate of NGC 3115 DW1 and its M/L ratio. Furthermore, we
rely in this section on the assumption that the system is not
influenced by the nearby giant S0 galaxy, an assumption that we will
question in Sec.~\ref{ln:discussion}.

We used two mass estimators which are extensively described and tested
by \citet{bahcall81} and \citet{heisler85}. We applied the Virial Mass
Estimator (VME) and the Projected Mass Estimator (PME) accounting for
different orbit characteristics of the globular clusters. Since the
alignment of our multi-slit masks is biased towards clusters of the
central and northern quadrant of the galaxy, our data is subject to
unknown systematic east-west rotation of the entire globular cluster
system. If the angular momentum vector points along the north-south
axis we will have under- or overestimated the mass depending on the
direction of rotation. Although, it is possible to provide a
lower-mass limit from our spatially constrained cluster sample by
varying the mean systemic velocity (see below), it is not possible to
correct completely for the unknown total rotation of the globular
cluster system (see also \ref{ln:rotdisp}).

\begin{figure}
\epsscale{1.0}
\plotone{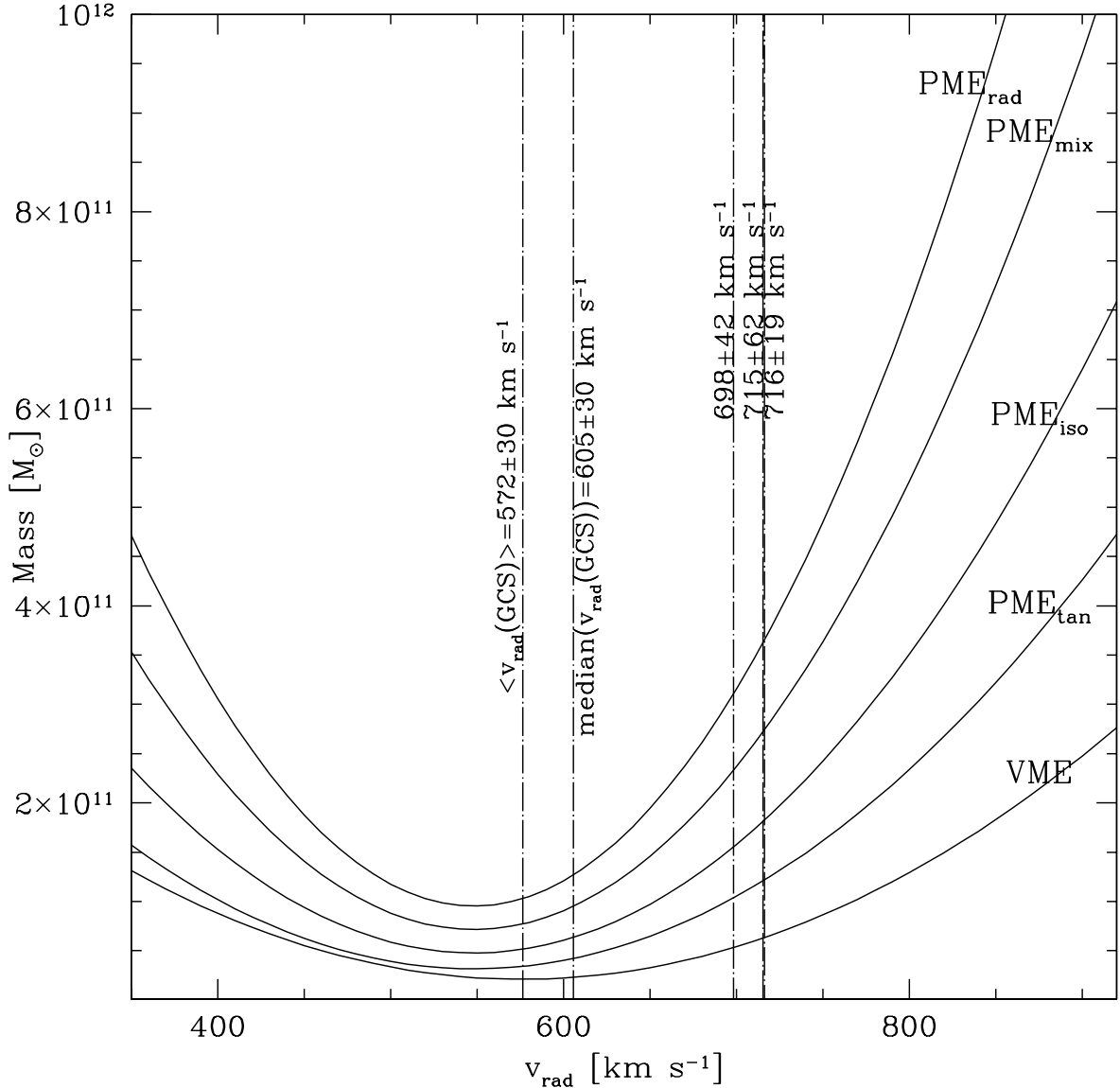}
\caption{Lower mass limit estimations with two different mass
estimators: PME (Projected Mass Estimator, allowing different orbit
characteristics of the GCS: rad -- radial, mix -- mixed, iso --
isotropic, and tan -- tangential orbits, see \citeauthor{heisler85}
1985 for details) and VME (Virial Mass Estimator). The measured mean
radial velocity of the globular cluster system is $v_{\rm rad}=572\pm
30$ km s$^{-1}$ while the median radial velocity was determined with
$v_{\rm rad}=605\pm30$ km s$^{-1}$. The curves show the variation of
the galaxy-mass estimate as a function of the mean system
velocity. Spectroscopy of the galaxy (NGC 3115 DW1) itself gives
radial velocities of $715\pm62$ km s$^{-1}$ \citep{RC3}, $716\pm19$ km
s$^{-1}$ \citep{peterson93}, and $698\pm42$ km s$^{-1}$
\citep{capaccioli93}. The errors of each mass estimator are given in
Table~\ref{tab:massestimate}.}
\label{ps:massestim}
\end{figure}

We let the radial velocity of NGC 3115 DW1 (which is assumed to be the
mean velocity of the GCS as well) vary over a wide range while the
measured radial velocities of the globular clusters remained fixed.
During each step of 10 km s$^{-1}$ we calculated the mass of the
galaxy with each mass estimator. The resulting plot (mass vs. $v_{\rm
  rad}$ of the GCS) is shown in Figure \ref{ps:massestim}. Table
\ref{tab:massestimate} shows the lower mass limits calculated with all
mass estimators.

At the highest measured radial velocity of NGC 3115 DW1 \citep[$v_{\rm
  rad}=716\pm19$ km s$^{-1}$,][]{peterson93} we obtain total masses in
the range $6.3\cdot10^{10}$M$_\odot$ to $3.6\cdot10^{11}$M$_\odot$.
The estimated masses for the measured mean radial velocity of the
globular cluster system ($v_{\rm rad}=572\pm30$ km s$^{-1}$) are in
the range $2.1\cdot10^{10}$M$_\odot$ to $1.0\cdot10^{11}$M$_\odot$.
The errors in Table \ref{tab:massestimate} are the intrinsic,
statistic, and systematic uncertainties of the mass-estimate. The
first is the uncertainty of the code itself \citep{bahcall81} while
the remaining are due to the limited sample size and the uncertain
distance. All masses are the total mass estimates within a
galactocentric radius of $r\leq189.4$\arcsec\ or $R\leq10.1$ kpc,
respectively. This is the projected radial distance of the outermost
globular cluster (L63).

Assuming an {\it isotropic} orbit distribution for globular clusters
in NGC 3115 DW1 the lower-mass limit for the galaxy of $M_{\rm
PME}=(4.8\pm2.3)\cdot10^{10}$M$_\odot$ seems rather large for a dwarf
elliptical. The absolute magnitude of $M_V=-17.7$ mag
\citep{durrell96} is high as well (more than a magnitude brighter than
M32 for example, and similar to NGC 4486B, although NGC 3115 DW1 has a
dissimilar structure to that of these low-L Es). The mean absolute
V-magnitude for nearby dwarf ellipticals is $<M_V>\approx-16.9$ mag
\citep{ferguson94}. Considering its mass and luminosity we address the
fact that NGC 3115 DW1 appears to be a transition-type galaxy between
luminous dEs and low-luminosity ellipticals in the discussion section.

\subsubsection{Radial Dependencies and Mass-to-Light Ratios}
\label{ln:radialdepend}

Limiting the data set of radial velocities to smaller radii we can
probe the radial mass dependencies in NGC 3115 DW1. Clearly,
statistical errors become important when we reduce the already small
data set by removing the outermost globular clusters. Nonetheless, we
calculate mass estimates for different radii (using the PME and
assuming isotropic orbits) since the kinematics of the two outer
clusters might be influenced by the nearby giant S0. The results are
summarized in Table \ref{tab:massrad}.

Combining the former findings with the photometry of
\citet{durrell96}, who found total magnitudes $V_{\rm T}=12.63\pm0.06$
mag and $B_{\rm T}=13.57\pm0.09$ mag, we estimate rough mass-to-light
ratios for different radii. Applying a King profile \citep{king62}
with the parameters $r_c=$14.4\arcsec\ and $c\sim$1.4
\citep{durrell96} we obtain at the outermost globular-cluster
projected radius $r=189.4$\arcsec\ ($R=10.1$ kpc) M/L$_V=52\pm25$ with
a $1\sigma$ uncertainty due to statistical sample-size uncertainties
and photometric errors. The systematic error due to distance
uncertainties of NGC 3115 DW1 is $^{+67}_{-32}$. Going inwards, the
M/L$_V$ ratio drops. At a projected radius of $r=56.3$\arcsec\ 
($R=3.0$ kpc) the mass-to-light ratio is M/L$_V=22\pm13$.

Our analysis can be expanded by combining our results with the M/L$_V$
measurements for the innermost part of NGC 3115 DW1.
\citet{peterson93} measure within $r\leq3$\arcsec\ (or $R\leq160$ pc)
of NGC 3115 DW1 a central velocity dispersion of $\sigma=20-30$ km
s$^{-1}$. They derive a M/L$_V$ of $3\pm2$ within $r\leq3$\arcsec. A
gradient of M/L$_V$ can therefore be traced outward from the center of
the galaxy, although the uncertainties are quite large. However, all
values fit well into the range given by \citet{ferguson94} for dwarf
elliptical galaxies (M/L$_V\approx5$ for Fornax up to M/L$_V\geq100$
for Ursa Minor).

\subsubsection{Rotation and Velocity Dispersion}
\label{ln:rotdisp}

Using the Maximum-Likelihood method of \citet{pryor_meylan93} we
measure a marginal net rotation of $v_{\rm rot}=75\pm70$ km s$^{-1}$
for the globular cluster system of NGC 3115 DW1. Note that to date no
dE was reported to show significant rotation of its stellar body
\citep{ferguson94}. The position angle of the rotation axis is
$\theta=90^{\rm o}\pm60^{\rm o}$ (poorly defined given the weak
rotation). This result may be of course biased by the incomplete
spatial coverage of our small sample. Correcting for the net rotation
we obtain a line-of-sight velocity dispersion of $\sigma=130\pm15$ km
s$^{-1}$ for the full sample of 7 clusters, and $\sigma=74\pm36$ km
s$^{-1}$ for the inner 5 clusters.

The major axis of the dE1,N galaxy of $\theta_{\rm gal}=100^{\rm o}
\pm10^{\rm o}$ \citep{durrell96} appears at face value nearly {\it
parallel} to the rotation axis of the GCS although the latter is not
well defined as mentioned above. Spectroscopy for more clusters with
better spatial coverage is certainly needed to establish the axis
alignment.

\subsection{Metallicity}

\subsubsection{Mean Metallicity of the Globular Cluster System}
\begin{figure}
\epsscale{1.0}
\plotone{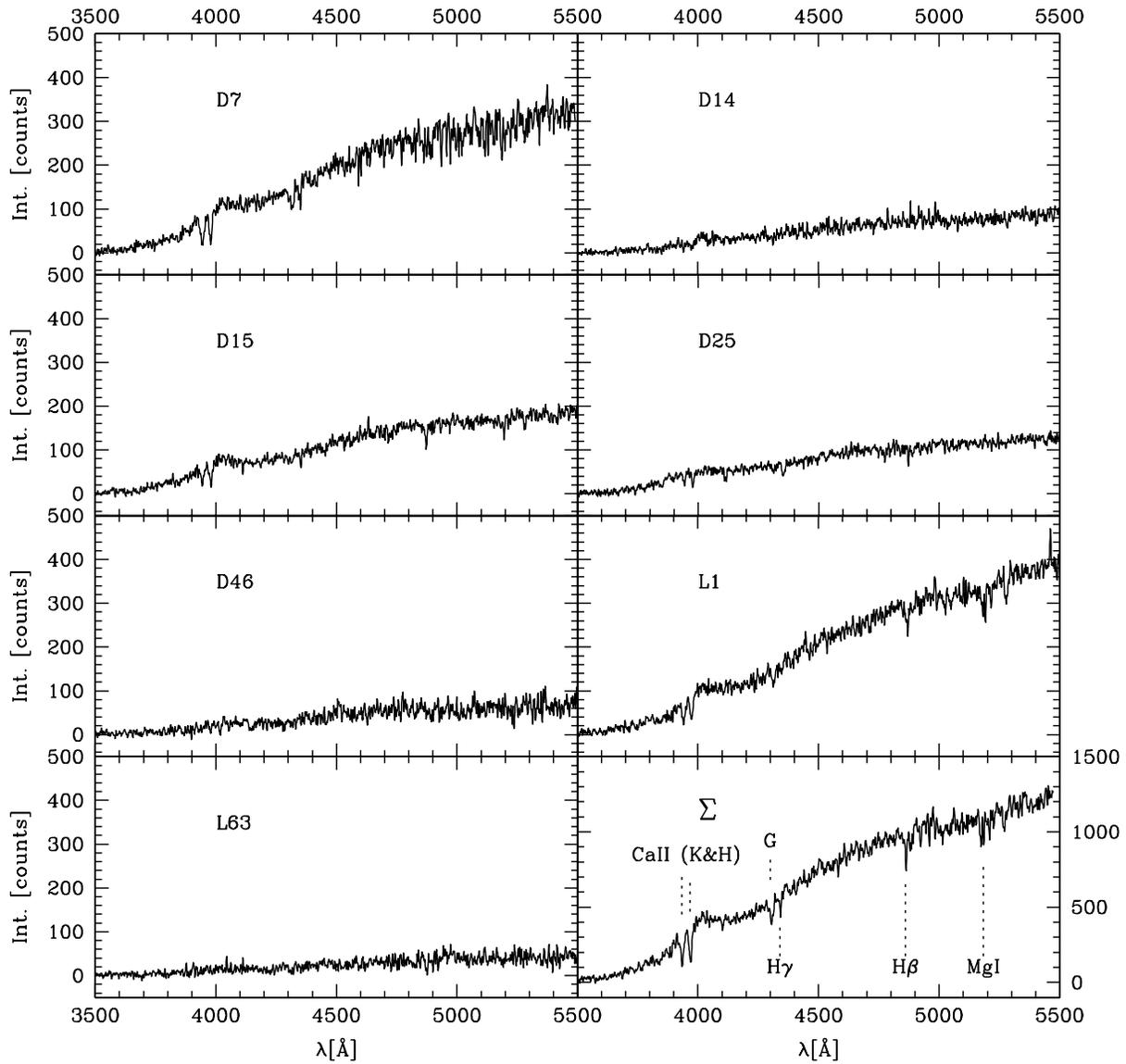}
\caption{Spectra of all 7 confirmed globular clusters. The S/N of the
spectra ranges between 2 and 5. The lower right panel shows the
combined spectrum of all other spectra. All spectra have been smoothed
with a 3-pix boxcar filter. Note the different scale of the lower
right panel (combined spectrum).}
\label{ps:spectra}
\end{figure}

The S/N of our spectra are insufficient to reliably establish
individual cluster metallicities. To measure a mean abundance for the
GCS we combined all the individual spectra into a high-S/N `mean'
spectrum. All individual globular cluster spectra as well as the
combined spectrum are shown in Figure \ref{ps:spectra}.

For abundance measurements we used the passband definitions of
\citet{brodie90} and the new Lick/IDS passband definitions of
\citet{trager98}. All abundances (both of single spectra and the mean
spectrum) are given in Table \ref{tab:abundmag}. \citet{brodie90}
calibrate single element abundances with [Fe/H] using a large sample
of Milky Way and M31 globular clusters. We used their calibration to
estimate the mean [Fe/H] for the entire GCS, based on the composite
spectrum. The mean metallicity of our globular cluster sample is
$\langle$[Fe/H]$\rangle_{\rm GCS}=-0.97\pm0.11$ dex. Exactly the same
value is obtained from the weighted mean of the individual
measurements (see Table \ref{tab:metalcomp}).

Overall, the mean metallicity does {\it not} follow the empirical
GCS-metallicity --- galaxy-luminosity relation \citep[see Fig.~5.7
in][]{ashman98}. NGC 3115 DW1 ($M_V=-17.7$ mag) falls in the
transition region between dwarfs and elliptical galaxies, whereas the
mean metallicity of the GCS falls in the range of metallicities found
in giant elliptical galaxies. According to this empirical relation NGC
3115 DW1 appears to be slightly too metal-rich for its luminosity.

\subsubsection{Abundance Ratios}

We compare the mean abundance ratios of the GCS with abundance ratios
of globular clusters in other galaxies. \citet{trager98} provide a
compilation of abundances of Milky Way and M31 globular clusters.
Abundances of NGC 1399 globular clusters were measured by
\citet{kisslerpatig98a}. Both data sets use the definitions of the
Lick/IDS system. In order to minimize the statistical noise of
abundance measurements we calculate a mean iron abundance
$\langle\mbox{Fe}\rangle$ and a mean metal abundance $[\mbox{MgFe}]$
\citep[see][for a detailed discussion]{gonzalez93}.

\begin{figure}
\epsscale{1.0}
\plotone{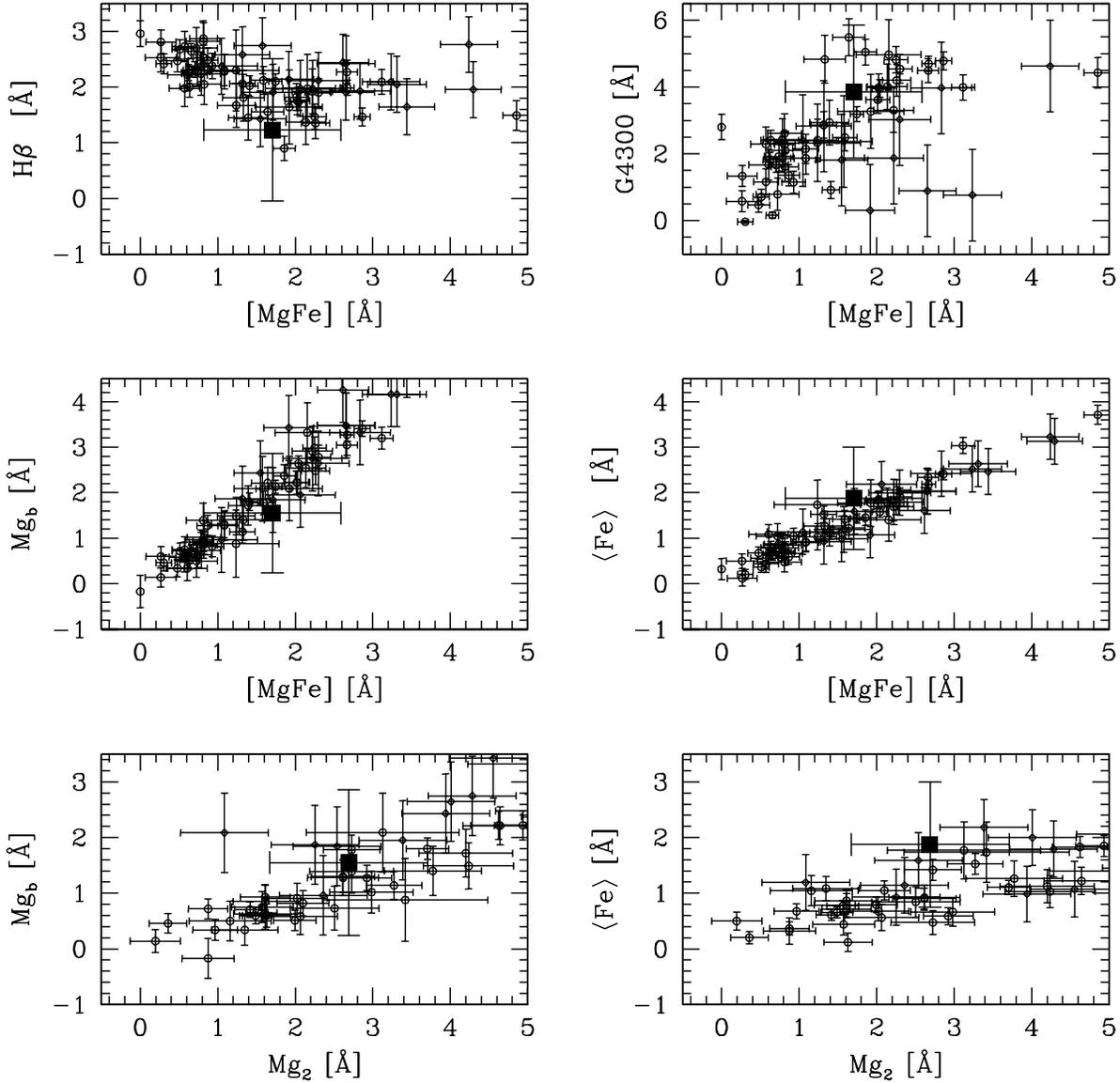}
\caption{Abundance ratios of the GCS in NGC 3115 DW (solid square)
compared to ratios of individual globular clusters in the Milky Way,
M31 \citep[open circles,][]{trager98}, and NGC 1399 \citep[open
diamonds,][]{kisslerpatig98a}. The abundances of the GCS in NGC 3115
DW1 were derived from a combined mean spectrum of 7 globular
clusters.}
\label{ps:metals}
\end{figure}

Figure \ref{ps:metals} shows abundance ratios for several dominant
elements. The upper four panels show abundance ratios relative to the
[MgFe] index. In the upper left panel the age-sensitive H$\beta$
abundance is seen to be in good agreement with Milky Way and M31 data,
although possibly at the lower (older) edge. The upper right panel
shows the G-band index compared with the [MgFe] index. The G-band is a
primary metallicity indicator \citep{brodie90}. The data show no
abundance anomalies.

The two middle panels of Figure \ref{ps:metals} could in principal be
used to examine the $\alpha$-element content of these globular
clusters. For the stellar light of (mostly brighter) ellipticals
\citet{worthey92} found an $\alpha$-element enhancement. The
[$\alpha$/Fe] enhancement is a very sensitive indicator of the star
formation rate in a galaxy. As $\alpha$-elements are preferentially
created in SNe type II, their enhancement indicates a violent star
formation and/or a top-heavy IMF. A depression, or normal values, of
[$\alpha$/Fe] would result from quiet star formation in which SNe Ia
dominate the enrichment processes. Given the relatively high mean
metallicity ($\langle$[Fe/H]$\rangle_{\rm GCS}=-0.97\pm0.11$ dex), a
normal $\alpha$-element ratio would suggest that these clusters formed
from enriched material either during epochs of quiet star formation,
or at the very beginning of a burst. Better data could lead to
interesting insights on this topic.

\section{Photometry}

\subsection{Spectroscopic check on photometry}

The main result from the photometry of \citet{durrell96} was that the
GCS in NGC 3115 DW1 is rich with a specific frequency of $S_{\rm
N}=4.9\pm1.9$ and a total globular cluster population $N_{\rm
GC}=59\pm23$. Our spectroscopy and the photometric data set of Durrell
et al.~have 22 objects in common (see Figure \ref{ps:fov} and Table
\ref{tab:specphot}). 6 of these 22 objects have projected radial
distances of $r\leq48$\arcsec\ ($R\leq2.6$ kpc). Durrell et
al.~consider the GCS to lie within this radius (mainly because the
surface over-density of objects disappears beyond it). We can confirm
4 of the 6 objects as bona-fide globular clusters (D7, D14, D15, and
D46). One object was found to be a background galaxy and another
cannot be identified either by radial velocity or by its spectrum.

Assuming that this sample of 6 objects is a statistically
representative sample of objects in the projected vicinity of NGC 3115
DW1 the upper limit of the probability of finding a globular cluster
within a radius of $r\leq48$\arcsec\ ($R\leq2.6$ kpc) around the
center of NGC 3115 DW1 is $f_{\rm GC}=5/6\approx0.83$ (the lower limit
is $f_{\rm GC}=4/6\approx0.67$; if we exclude the non-identified
object). This is in good agreement with the findings of
\citet{durrell96}, although the statistical significance is very low.
Durrell et al.~measure the contaminating surface density of background
objects to be $\sigma=6.4\pm1.9$ arcmin$^{-2}$. Within their radial
limit of $r\leq48$\arcsec\ there are a total of $\sim13$ background
objects. For a total population of $N_{\rm GC}=59\pm23$, the
probability of picking a globular cluster within $r\leq48$\arcsec\ 
(and at the magnitude limit of the photometry of Durrell et al.) is
$f_{\rm GC}=59/(13+59)\approx0.82$. Our spectroscopic results suggest
that there is no need to make any correction to the values for
specific frequency and total globular cluster population size derived
from photometry.

\citet{miller98} measure specific frequencies for dwarf elliptical
galaxies in the Virgo and Fornax cluster and find a $\log(S_{\rm
  N})-M_V$ relations for the nucleated dwarfs. NGC 3115 DW1 has a
higher specific frequency ($S_{\rm N}=4.9\pm1.9$) than the $S_{\rm
  N}\approx2.2$ derived from Miller's et al.~relation for group and
cluster dE,N galaxies.

\begin{figure}
\epsscale{1.0}
\plotone{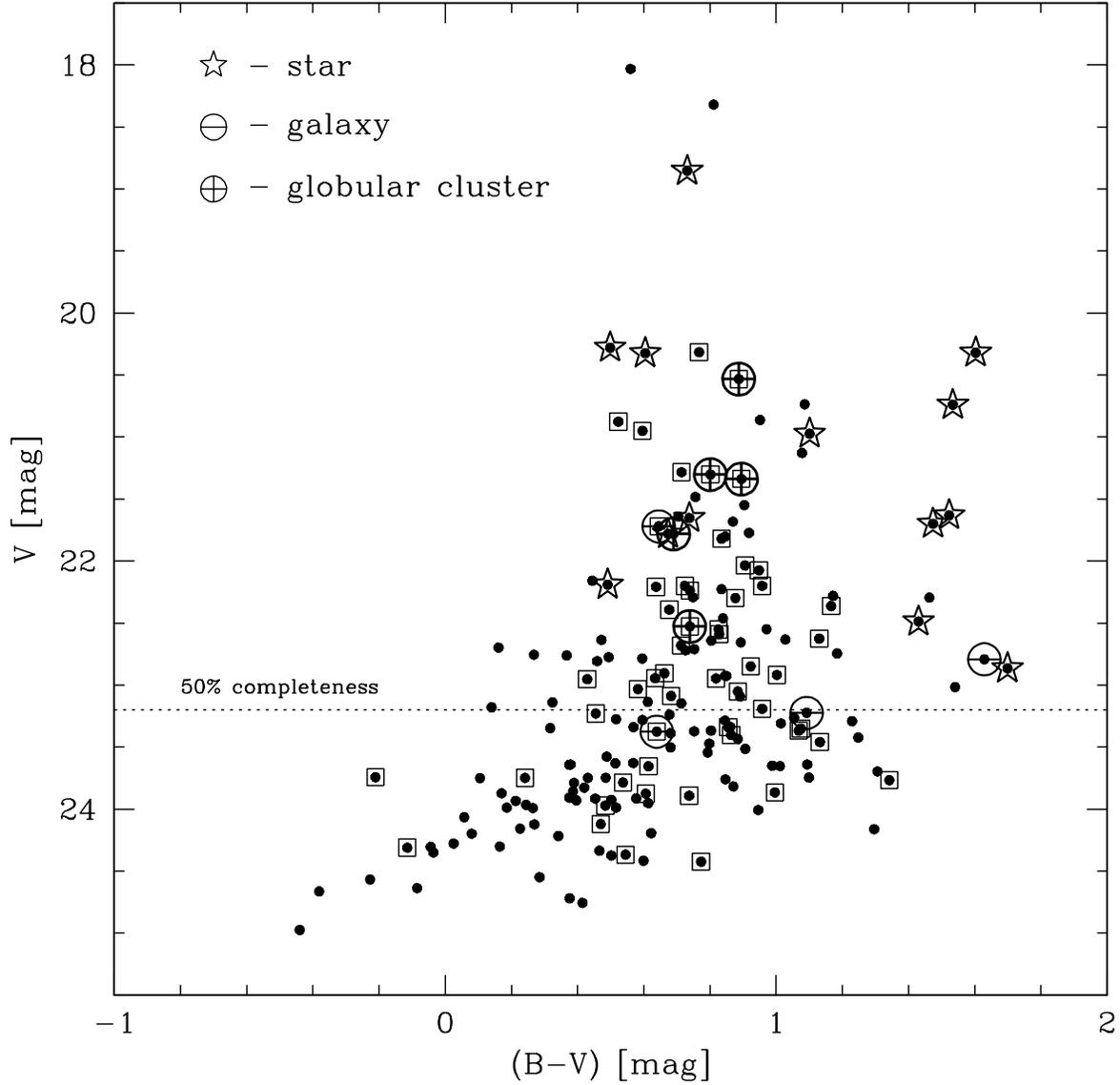}
\caption{Color magnitude diagram of objects in the vicinity of NGC
3115 DW1. Dots are objects with photometry by \citet{durrell96}.  Open
squares mark objects with $r\leq48$\arcsec\ ($R\leq2.6$ kpc) projected
distance to the galaxy center. Open stars are spectroscopically
verified foreground stars, $\ominus$ mark background galaxies, and
$\oplus$ are bona-fide globular clusters.  There are only 5
spectroscopically confirmed globular clusters with $B,V$
photometry. Two globular clusters (L1 and L63) are not covered by the
field-of-view of the \citeauthor{durrell96} photometry (see
Fig.~\ref{ps:fov}). The dotted line indicates the photometric
50\%-completeness limit in V.}
\label{ps:cmd}
\end{figure}

Figure \ref{ps:cmd} shows the CMD of objects with $r\leq48$\arcsec\ 
($R\leq2.6$ kpc) which have been marked by open squares. Objects with
photometry by Durrell et al., spectroscopically identified foreground
stars, background galaxies, and globular clusters are indicated. We
determined the mean color of 5 globular clusters (2 spectroscopically
confirmed globular clusters are not included in the photometric
sample) in this CMD with Maximum-Likelihood techniques. We obtained
$\langle(B-V)\rangle_{\rm GCS}=0.82\pm0.04$ mag with a dispersion of
$\sigma(B-V)_{\rm GCS}=0.06\pm0.04$ mag. Durrell et al. found
$\langle(B-V)\rangle=0.74\pm0.03$ mag, $\sigma(B-V)=0.13$ mag for the
total sample i.e.~corresponding to a lower mean metallicity. Our
subset seems to be slightly biased towards metal-rich objects.

\subsection{Comparison of photometrically- and
  spectroscopically-derived metallicities} 

In order to constrain the significance of photo\-metrically-derived
metallicities we transform the $(B-V)$ color into a [Fe/H]-metallicity
and compare it with the findings of our abundance measurements. For
this purpose we use the relation of \citet{couture90}
\begin{equation}
  \label{eq:colormetallin}
  \mbox{[Fe/H]}=5.0\cdot(B-V)_{\rm o}-4.86
\end{equation}
with [Fe/H] being the independent parameter during the calibration of
the equation. The application of equation \ref{eq:colormetallin} to
all de-reddened \citep[E$_{(B-V)}=0.052$ mag,][]{schlegel98}
globular-cluster colors in our NGC 3115 DW1 sample leads to
photometrically derived metallicities which can be compared with the
[Fe/H] values from spectroscopy. All data are summarized in Table
\ref{tab:metalcomp}. The resulting weighted mean metallicity is
$\langle$[Fe/H]$\rangle_{\rm GCS}=-0.93\pm0.11$ dex (for the 5
globular clusters) with a dispersion of $\sigma($[Fe/H]$)_{\rm
GCS}=0.41\pm0.20$ dex in good agreement with the values derived from
spectroscopy. The uncertainty results from the photometric error of
the color only. No transformation uncertainty was included.

\subsection{Globular Cluster Sizes}

We matched two spectroscopically-confirmed globular clusters, D15 and
D25 (see Figure \ref{ps:fov} and Table \ref{tab:specphot}), in the HST
images taken from the archive (both on WF chips). At the distance of
NGC 3115 DW1 the WF chips resolve globular clusters with core radii of
$r_c\sim$5 pc. Only 14\% of Milky-Way globular clusters show core
radii larger that 5 pc \citep{harris99}. However, we can use the HST
data to derive upper limits for the globular cluster sizes.  Only
images taken through the F814W filter were used because of their
higher S/N.

The radial source profile $I(r)$ (i.e. the PSF of the final image) is
a convolution of the object profile $O(r)$ with the telescope PSF
$T(r)$ and an additive noise term $R(r)$ ($r$ being the radial
distance from the center of the profile);
\begin{equation}
  \label{eq:deconv}
  I(r)=\int\limits_{r_{\rm min}}^{r_{\rm max}}O(s)\cdot T(r,s)\; ds + R(r).
\end{equation}
To calculate the telescope-PSF profile $T(r)$ we used the TinyTim v4.4
code by \citet{krist97} which gives a semi-analytic estimation of the
HST-PSF for each chip, each chip position, and each filter. We adopted
a King profile \citep{king62} for $O(r)$ which appears to be a good
fit to globular-cluster radial profiles in Milky Way
\citep[e.g.][]{trager95} and extragalactic systems
(\citeauthor{grillmair96} 1996 in M31, \citeauthor{elson85} 1985 in
LMC, \citeauthor{kundu98} 1998 in NGC 3115, \citeauthor{kundu99} 1999
in M87, and \citeauthor{puzia99} 1999 in NGC 4472). From an analysis
of Milky Way globular clusters \citep{harris99} we chose $c\equiv\log
(r_t/r_c)=1.5$ as a concentration parameter.

In equation \ref{eq:deconv} we neglect the additive noise term since
our size-estimation errors are dominated by the convolution of the
poorly-defined charge-diffusion matrix with the optical HST-PSF
\citep[see][for a detailed discussion]{krist97}. Note that the
charge-diffusion smears 25\% of the infalling light of the central
pixel among its neighbors. For consistency with other work we continue
to use this convolution throughout our analysis despite the fact that
the diffusion correction has been derived only for the F555W filter
and is thought to be wavelength dependent.

\begin{figure}
\epsscale{0.6}
\plotone{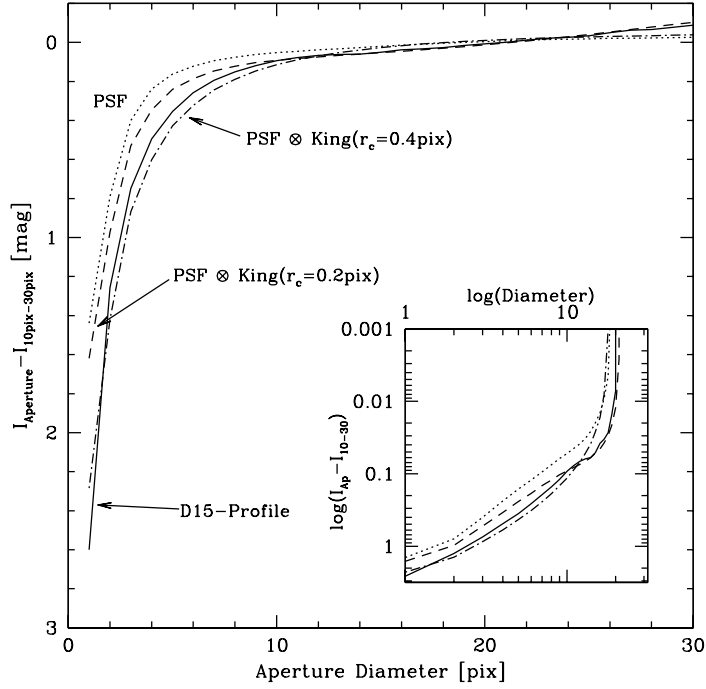}
\plotone{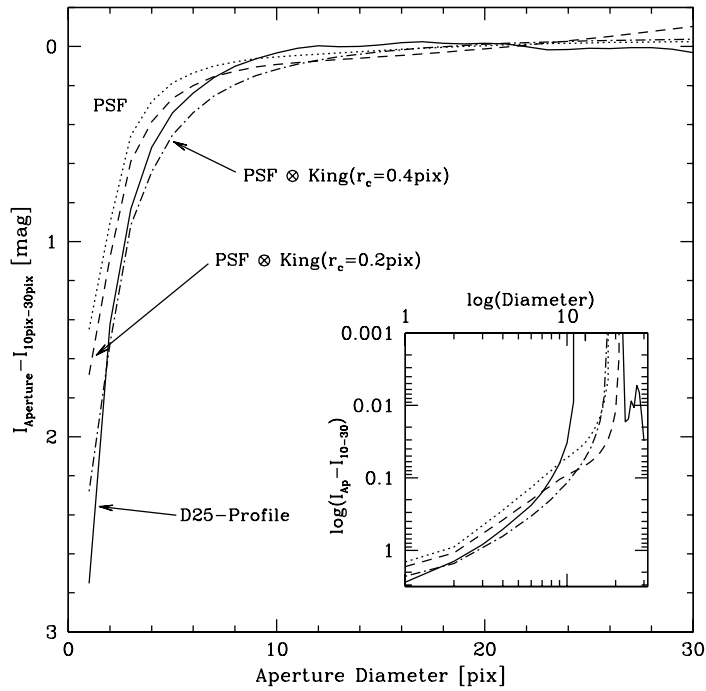}
\caption{Radial intensity profiles for globular cluster D15 (upper
panel) and D25 (lower panel). Two synthetic profiles are shown which
were created by a convolution of two King profiles with core radii
$r=0.2$ pix ($R=0.5$ pc, dashed curve) and $r=0.4$ pix ($R=1.0$ pc,
dot-dashed curve) and the corresponding HST-PSF. Also shown is a
point-source HST-PSF (dotted curve). The inlays show the same curves
in log-log scale.}
\label{ps:profile}
\end{figure}

Five King profiles were generated with core radii in the range
$r_c=0.1-0.5$ pix ($R_c=0.5-2.7$ pc). In addition, we generated
HST-PSFs for both our identified globular clusters using individual
specifications (e.g. filter, chip, chip position). Both HST-PSFs were
convolved with all the King profiles. Aperture photometry was applied
to all generated profiles and both globular clusters on the HST
images. For this purpose we used SExtractor \citep{bertin96} and
measured magnitudes in 30 apertures with diameters in the range 1--30
pix. All magnitudes were normalized to the average aperture magnitudes
in the range 10--30 pix, i.e.~$I(r)=I_o(r)-\langle I\rangle_{\rm
  10-30\: pix}$. Figure \ref{ps:profile} shows the profiles of both
globular clusters, a raw HST-PSF profile, and two of the convolution
profiles with core radii of $r_c=0.2$ and 0.4 pix.

Both globular cluster profiles deviate significantly from the raw
HST-PSF profile which indicates that both D15 and D25 are resolved.
The S/N of the F814W image drops to 1 at an aperture diameter of 6
pix. For larger apertures there is insufficient signal to detect any
deviations from a raw HST-PSF. For smaller apertures both globular
cluster profiles lie between King profiles of core radii 0.2 pix and
0.4 pix. We deduce an upper limit for both globular cluster sizes of
$r_c= 2.1^{+0.9}_{-0.4}$ pc at an adopted distance of NGC 3115 DW1 of
$d= 11^{+5.0}_{-2.3}$ Mpc \citep{durrell96}. 27\% of Milky-Way
globular clusters have core radii larger than 2.1 pc \citep{harris99}
and 10\% have sizes in the range defined by the errors of the NGC 3115
DW1 clusters.

This upper limit compares well with the results of \citet{kundu98} and
\citet{kundu99}. Using HST photometry, these authors find typical
half-light radii\footnote{The half-light radius is comparable with the
  core radius of a King profile.} of $r_h=2.0\pm0.1$ pc and
$r_h\approx 2.5$ pc for globular clusters in NGC 3115 and M87,
respectively.

\section{Discussion}
\label{ln:discussion}

In section \ref{ln:rotdisp}, we derived a high globular cluster
velocity dispersion, and thus a high galaxy mass, when we included the
two outermost globular clusters. The high mass is not unexpected given
the bright absolute magnitude of NGC 3115 DW1. Based on its
$M_B=-16.8$ mag \citep{durrell96}, NGC 3115 DW1 falls in the
transition region between dwarfs and ellipticals in the
mass-luminosity relation of \citet{dekel86} (see Fig.~3 therein). Its
high mass ($M_{\rm PME}=(4.8\pm2.3)\cdot10^{10}$M$_\odot$) and the
high velocity dispersion ($\sigma=130\pm15$ km s$^{-1}$, see below)
are more consistent with an elliptical galaxy. We therefore discuss
whether the two outermost clusters could be in the process of being
stripped by the nearby giant S0 galaxy NGC 3115.

\subsection{Possible Stripping?}

\begin{figure}
\epsscale{1.0}
\plotone{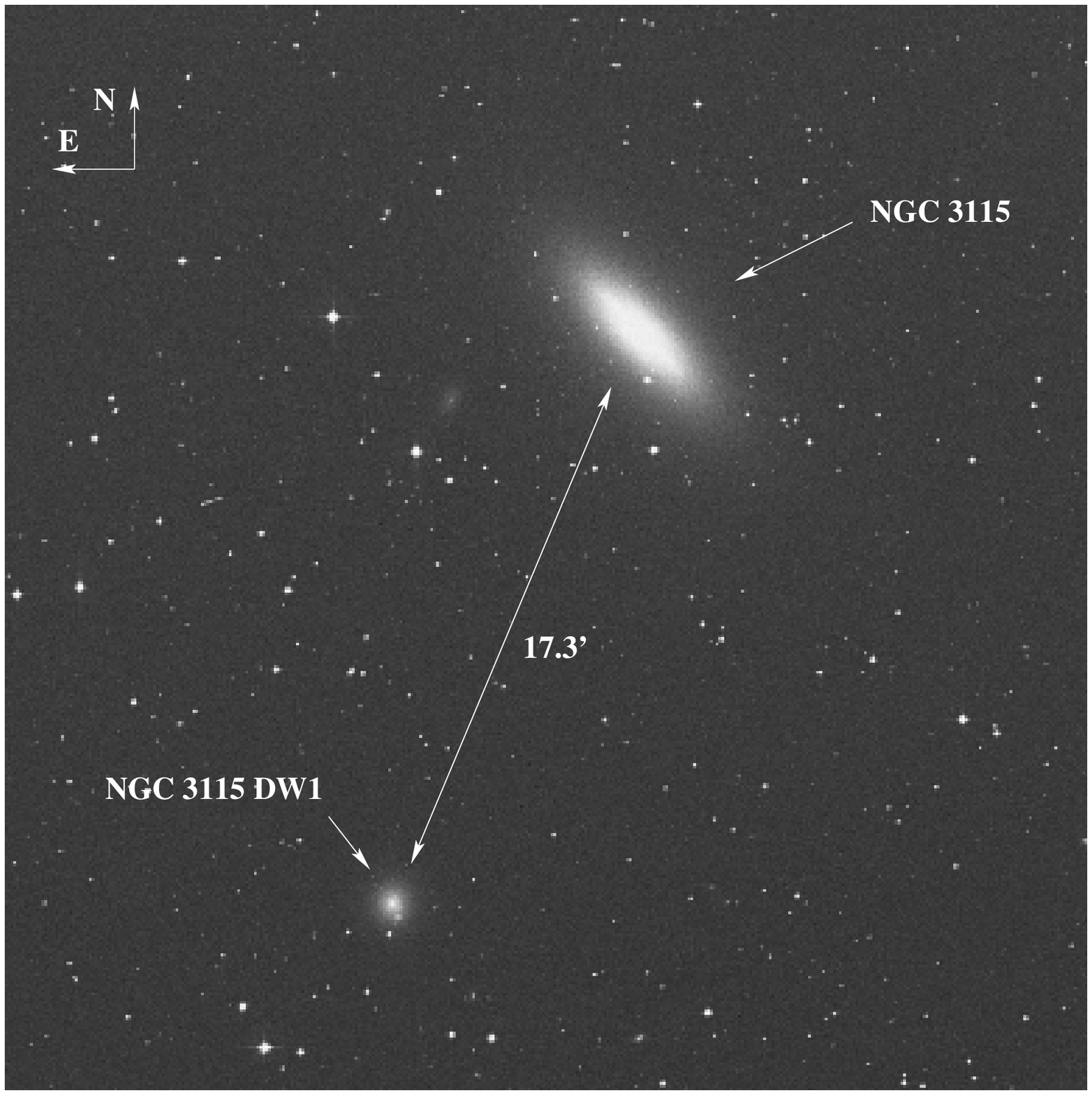}
\caption{Relative positions of the dwarf elliptical galaxy NGC 3115
DW1 and the S0 galaxy NGC 3115. The projected distance between these
two galaxies is 17.3\arcmin\ (55 kpc). The image was taken from the
Digitized Sky Survey. The size is 30\arcmin\ $\times$ 30\arcmin .}
\label{ps:localgroup}
\end{figure}

Figure \ref{ps:fov} and Table \ref{tab:specphot} show that 2 (L1 and
L63) of the 7 globular clusters have significantly larger projected
radii, i.e. 161.4\arcsec\ ($8.6$ kpc, L1) and 189.4\arcsec\ ($10.1$
kpc, L63), than the ``inner'' ($r\leq56.3$\arcsec $=3$ kpc) globular
clusters. These large projected distances from NGC 3115 DW1 could be
due to stripping by the nearby S0 galaxy NGC 3115. Figure
\ref{ps:localgroup} shows the relative positions of NGC 3115 DW1 and
NGC 3115. The projected distance between the two galaxies is
17.3\arcmin\ which corresponds to 55 kpc at the distance of
$d\approx$11 Mpc. The mean radial velocities of L1 and L63 are $v_{\rm
  rad}=420\pm29$ km s$^{-1}$ and $v_{\rm rad}=605\pm74$ km s$^{-1}$,
respectively. Only L1 shows a significant deviation from the systemic
velocity of NGC 3115 DW1 \citep[$v_{\rm rad}=698\pm74$ km
s$^{-1}$,][]{capaccioli93} and NGC 3115 \citep[$v_{\rm rad}=663\pm6$
km s$^{-1}$,][]{capaccioli93}.

We expect no contamination from globular clusters of the nearby galaxy
NGC 3115. \citet{kavelaars97} found the surface density over-abundance
of globular clusters around NGC 3115 (power-law index of radial
distribution $\alpha=-1.8\pm0.5$) disappearing at 6\arcmin\ radius
from the center of NGC 3115 (at a photometric limit of $V=23.5$ mag).
The globular clusters L1 and L63 have a radial distance to NGC 3115 of
$\approx14$\arcmin . The projected GC surface density of the GCS of
NGC 3115 at the position of these two clusters is $<0.01$
arcmin$^{-2}$. The extrapolated GC surface density of NGC 3115 DW1 at
this position lies between 0.2 and 6.9 GCs arcmin$^{-2}$, given the
large uncertainties on the density profile. As the numbers are too
small (we only found 2 clusters to the north and 0 to the south) it
cannot statistically be concluded whether the two globular clusters
found in the northern field are chance detections or a statistically
significant overabundance.

Assuming that both galaxies are roughly at the same distance, we can
estimate the dwarf galaxy's gravitational potential and the ratio of
potentials of NGC 3115 DW1 and NGC 3115. Both globular clusters are at
about 1/5 of the distance separating NGC 3115 DW1 and NGC 3115. As a
rough estimate, we assume that NGC 3115 DW1 and NGC 3115 have similar
M/L$_V$. In this case, the ratio of their $M_V$'s would imply that NGC
3115 has a mass 10 times larger than NGC 3115 DW1. Hence, the
gravitational potentials are comparable at the projected position of
the distant globular clusters (L1 and L63). Since the mass of NGC 3115
is likely to be higher than the adopted value (assuming an extended
dark matter halo) the motion of both globular clusters is no longer
dominated by the gravitational potential of NGC 3115 DW1 alone. Both
clusters could then be considered as intergalactic globular clusters.

Note that stripping of globular clusters appears to be common among
interacting galaxies. \citet{dacosta95} show that four globular
clusters of the Sagittarius dSph are in the process of being stripped
by the Milky Way and are being added to its globular cluster system.
Other studies have indicated that stripping may be important in galaxy
clusters (e.g. in the Fornax cluster \citeauthor{kisslerpatig99} 1999,
\citeauthor{hilker99} 1999). However, there are no other (optical)
hints of interaction from NGC 3115 DW1's stellar light.
\citet{durrell96} found the isophotes to be consistent with little or
no tidal disruption out to a projected radius of 60\arcsec\
(corresponding to 3.2 kpc) where their photometric errors start to
dominate.

A simple test for the stripping hypothesis would be a wide-field study
of the system in order to rule out (spectroscopically) the presence of
any similar clusters around NGC 3115 DW1.

\subsection{The Expected Velocity Dispersion}

A look at the fundamental plane of dwarf elliptical galaxies
\citep[e.g.][]{peterson93} shows that NGC 3115 DW1 fits reasonably
well into the relation for dwarf and giant elliptical galaxies, under
their assumption of $M_V=-16.7$ mag. Adopting the absolute magnitude
of $M_V=-17.7$ mag \citep{durrell96b} the galaxy falls slightly off
the relation and would imply a higher velocity dispersion than
measured in the central 3\arcsec . With the measured velocity
dispersion of $\sigma=74\pm36$ km s$^{-1}$ for the 5 globular clusters
inside $r<56.3$\arcsec\ ($R<3$ kpc) we obtain from the
fundamental-plane relation of \citet{peterson93} an absolute magnitude
of $M_V=-18.0\pm0.5$ mag. The measured velocity dispersion for the
total sample of 7 clusters inside $r<189.4$\arcsec\ ($R<10.1$ kpc) of
$\sigma=130\pm15$ km s$^{-1}$, would correspond to far brighter
absolute magnitude ($M_V=-19.5\pm0.5$ mag) than the measured
$M_V=-17.7$ mag. This discrepancy can be explained by a close
encounter and subsequent stripping of the dwarf galaxy's halo by the
nearby S0 galaxy NGC 3115. Stripping of outer halo regions might well
have introduced violent perturbations and led to an enhanced velocity
dispersion of the halo region (which is traced by the globular
clusters). The relaxation time of such a system far exceeds the Hubble
time \citep{binney94} and therefore it is not possible to reject this
scenario just from considerations of dynamical timescales.

Alternatively, a high velocity dispersion in the outskirts of a galaxy
could be due to a dark-matter dominated massive halo. The outer parts
of a number of lower-luminosity Local Group galaxies are known to be
dominated by dark matter \citep[e.g.][]{mateo98}. This picture could
explain the fact that we measure an uncommonly high mass for a dwarf
elliptical (see sec.~\ref{ln:massestimate}) at a projected radius of
189.4\arcsec\ ($10.1$ kpc).

We cannot discriminate between the above possibilities at this point.

\section{Summary}

Using LRIS multi-slit spectra we confirm, on the basis of their radial
velocities, 7 of the 46 objects in our spectroscopic sample as
bona-fide globular clusters associated with the bright ($M_V=-17.7$
mag) dE1,N galaxy, NGC 3115 DW1.

We verify the findings of \citet{durrell96} (within a projected radius
of $r\leq48$\arcsec\, corresponding to $R\leq2.6$ kpc) who derived the
specific frequency $S_{\rm N}=4.9\pm1.9$ and a total globular cluster
population size $N_{\rm GC}=59\pm23$. The spectroscopic verification
of foreground and background contamination indicates that no revision
of these results is necessary. NGC 3115 DW1 remains a dE,N galaxy with
a somewhat high $S_{\rm N}$ value.

A mass estimate using the Projected Mass Estimator (PME) for isotropic
globular cluster orbits yields a total galaxy mass of $M_{\rm
gal}=(4.8\pm2.3)\cdot10^{10}M_\odot$ (with the error being the
internal uncertainty of the mass estimation code) inside a radius
$r\leq189.4$\arcsec\ ($R\leq10.1$ kpc) and $M_{\rm
gal}=(1.8\pm1.0)\cdot10^{10}M_\odot$ inside $r\leq56.3$\arcsec\
($R\leq3.0$ kpc). This estimate is a lower mass limit (see
\ref{ln:massestimate}) and assumes that the outer globular clusters
are not influenced by the nearby giant S0 NGC 3115. Using two mass
estimators (i.e.~PME and VME, see Sect.~\ref{ln:massestimate}) and
various assumptions for the systemic velocity and the cluster orbits,
we derive masses between $2 \cdot 10^{10}$M$_\odot$ and
$4\cdot10^{11}$ M$_\odot$. The mass increases with radius (see Table
\ref{tab:massrad}). Inside R$<160$ pc the mass-to-light ratio was
found to be M/L$_V=3\pm2$ \citep{peterson93} and increases with
radius, leading to M/L$_V=52\pm25$ at $\sim$ 10 kpc (using the PME and
assuming isotropic orbits), compatible with dark matter dominated
outer regions. However, we cannot at present exclude the possibility
that the high velocity dispersion is due to stripping of the two outer
clusters by the nearby giant companion.

A kinematic analysis shows that the globular cluster system has a
marginal net rotation of $v_{\rm rot}=75\pm70$ km s$^{-1}$ with a
position angle of the rotation axis $\theta=90^{\rm o}\pm60^{\rm
o}$. Subtracting the net rotation we find a line-of-sight velocity
dispersion of the globular cluster system of $\sigma=130\pm15$ km
s$^{-1}$ for the total sample of 7 globular clusters and
$\sigma=74\pm36$ km s$^{-1}$ for the inner 5 clusters (see
Sect.~\ref{ln:rotdisp}).

We measure mean abundances (using Lick/IDS passband definitions) from
a combined mean spectrum of all 7 globular clusters and derive a mean
GCS metallicity of $\langle$[Fe/H]$\rangle_{\rm GCS}=-0.97\pm0.11$
dex. All abundance ratios appear similar to the ones measured in Milky
Way, M31 and NGC 1399 globular clusters.

The mean color of the spectroscopically confirmed globular clusters is
$\langle(B-V)\rangle_{\rm GCS}=0.82\pm0.04$ mag with a dispersion
$\sigma(B-V)_{\rm GCS}=0.06\pm0.04$ mag.

Applying the color-metallicity calibration of \citet{couture90} we
obtain a photometric mean metallicity $\langle$[Fe/H]$\rangle_{\rm
  GCS}=-0.93\pm0.11$ dex with a dispersion of
$\sigma(\mbox{[Fe/H]})_{\rm GCS}=0.41\pm0.20$ dex (the error being the
photometric uncertainty).

For two globular clusters (L1 and L63) with HST photometry we derive
upper limits for their core radii. These were found to be
$r_c=2.1^{+0.9}_{-0.4}$ pc.

\acknowledgments 

We would like to thank Patrick Durrell for providing his ground-based
photometry in electronic form. We also thank Duncan Forbes for his
help during the observation.

Some of the data presented herein were obtained at the W.M. Keck
Observatory, which is operated as a scientific partnership among the
California Institute of Technology, the University of California and
the National Aeronautics and Space Administration. The Observatory was
made possible by the generous financial support of the W.M. Keck
Foundation.

This work was supported by National Science Foundation grant number
AST990732 and Faculty Research funds of the University of California,
Santa Cruz.

\clearpage
\onecolumn 
\begin{deluxetable}{rr rl rl rl rlc}
\tabletypesize{\scriptsize}
\tablecaption{Radial Velocities from Cross-Correlation and individual
  Wavelength Shifts \label{tab:rvs}}
\tablewidth{0pt}
\tablehead{
\colhead{slit No.\tablenotemark{a}} & 
\colhead{code\tablenotemark{b}} & 
\multicolumn{2}{c}{$V_{\rm rad}^{158-213}$} &
\multicolumn{2}{c}{$V_{\rm rad}^{225-280}$} &
\multicolumn{2}{c}{$V_{\rm rad}^{\rm lines}$} & 
\multicolumn{2}{c}{$<V_{\rm rad}>$} &
\colhead{cat.\tablenotemark{c}}
}
\startdata
\multicolumn{2}{c}{} & \multicolumn{9}{c}{all radial velocities in km
  s$^{-1}$}\\\cline{3-10}\\
\multicolumn{11}{c}{mask A}\\\tableline
\multicolumn{11}{c}{}\\
 1&L15& 70492$\pm$& 336 &  70386$\pm$& 168 &  \nodata &          &  70407$\pm$ &  106 &$\ominus$\\
 2& L3&    63$\pm$& 100 &     81$\pm$& 136 &  \nodata &          &     69$\pm$ &   57 &$\star$\\
 3&L27& 25223$\pm$& 242 &  25242$\pm$& 189 &  \nodata &          &  25235$\pm$ &  105 &$\ominus$\\
 4&L11&    66$\pm$& 113 &    105$\pm$& 120 &  \nodata &          &     84$\pm$ &   58 &$\star$\\
 5&L18&   194$\pm$&  74 &    208$\pm$&  97 &  \nodata &          &    199$\pm$ &   42 &$\star$\\
 6& L7& $-$50$\pm$& 195 &  $-$65$\pm$& 141 &  \nodata &          &  $-$60$\pm$ &   81 &$\star$\\
 7&L16&$-$211$\pm$& 178 & $-$183$\pm$& 162 &  \nodata &          & $-$196$\pm$ &   85 &$\star$/?\\
 8& D8&    67$\pm$& 127 &    101$\pm$& 111 &  \nodata &          &     86$\pm$ &   59 &$\star$\\
 9&D21& \nodata   &     &  \nodata   &     &  \nodata &          &  \nodata    &      &??\\
10&D10&   100$\pm$&  85 &    118$\pm$&  73 &  \nodata &          &    110$\pm$ &   39 &$\star$\\
11&D58&  1864$\pm$& 153 &  23692$\pm$& 225 &  \nodata &          &   8766$\pm$ &   89 &$\ominus$/??\\
12&D73&  4987$\pm$& 307 &   5030$\pm$& 225 &  \nodata &          &   5015$\pm$ &  128 &$\ominus$/??\\
13&D20& 22542$\pm$& 226 &  22576$\pm$& 192 &  \nodata &          &  22562$\pm$ &  103 &$\ominus$\\
14&D24&   160$\pm$& 224 &    154$\pm$& 223 &  \nodata &          &    157$\pm$ &  112 &$\star$\\
15&D14&   455$\pm$& 156 &    477$\pm$& 149 &  478$\pm$&      190 &    468$\pm$ &   71 &$\oplus$/?\\
16& D6&    17$\pm$&  61 &     32$\pm$&  66 &  \nodata &          &     24$\pm$ &   32 &$\star$\\
17& D1&   143$\pm$&  73 &    152$\pm$&  83 &  \nodata &          &    147$\pm$ &   39 &$\star$\\
18& D5&   124$\pm$&  70 &    139$\pm$&  73 &  \nodata &          &    131$\pm$ &   36 &$\star$\\
19&L37& 24531$\pm$& 180 &  24614$\pm$& 243 &  \nodata &          &  24560$\pm$ &  102 &$\ominus$/??\\
20&L44& 22979$\pm$& 210 &  23020$\pm$& 223 &  \nodata &          &  22998$\pm$ &  108 &$\ominus$/??\\
21&L14&    34$\pm$& 111 &     67$\pm$& 103 &  \nodata &          &     52$\pm$ &   53 &$\star$\\
22&L51&    26$\pm$& 187 &     71$\pm$& 196 &  \nodata &          &     47$\pm$ &   96 &$\star$\\
23& L1&   386$\pm$&  76 &    402$\pm$&  79 &  453$\pm$&       44 &    420$\pm$ &   29 &$\oplus$\\
24&L63&   545$\pm$& 227 &    506$\pm$& 222 &  668$\pm$&       99 &    605$\pm$ &   74 &$\oplus$/?\\
\multicolumn{11}{c}{}\\
\multicolumn{11}{c}{mask B}\\\tableline
\multicolumn{11}{c}{}\\
 1& L8& 99434$\pm$& 137 &  99453$\pm$& 125 &  \nodata &          &  99444$\pm$ &   65 &$\ominus$\\
 2& L4&    55$\pm$& 136 &     45$\pm$& 109 &  \nodata &          &     49$\pm$ &   60 &$\star$\\
 3& L6&    \nodata&     &     \nodata&     &  \nodata &          &     \nodata &      &??\\ 
 4&L23&$-$337$\pm$& 227 & $-$259$\pm$& 201 &  \nodata &          & $-$293$\pm$ &  106 &$\star$/?\\
 5& D9&     9$\pm$& 128 &     36$\pm$& 117 &  \nodata &          &     24$\pm$ &   61 &$\star$\\
 6&D36& $-$23$\pm$& 152 &     13$\pm$& 140 &  \nodata &          &   $-$4$\pm$ &   73 &$\star$\\
 7&D26& $-2599\pm$& 241 &$-$2545$\pm$& 201 &  \nodata &          &$-$2567$\pm$ &  109 &$\star$/??\\
 8& D2& $-$37$\pm$& 111 &     24$\pm$& 129 &  \nodata &          &  $-$11$\pm$ &   59 &$\star$\\
 9&D25&   540$\pm$& 146 &    512$\pm$& 158 &  540$\pm$&      100 &    532$\pm$ &   60 &$\oplus$\\
10&D46&   628$\pm$& 215 &    619$\pm$& 208 &  \nodata &          &    623$\pm$ &  106 &$\oplus$/??\\
11& D3& $-$22$\pm$& 113 &     12$\pm$& 112 &  \nodata &          &   $-$5$\pm$ &   56 &$\star$\\
12&D15&   677$\pm$&  67 &    692$\pm$&  72 &  672$\pm$&       62 &    681$\pm$ &   30 &$\oplus$\\
13& D7&   685$\pm$&  70 &    706$\pm$&  74 &  710$\pm$&       34 &    703$\pm$ &   25 &$\oplus$\\
14&D42& 42007$\pm$& 264 &  70422$\pm$& 231 &  \nodata &          &  58100$\pm$ &  123 &$\ominus$/?\\
15&L36& 59943$\pm$& 215 &   3899$\pm$& 176 &  \nodata &          &  26386$\pm$ &   96 &$\ominus$/??\\
16&D41&   262$\pm$& 206 &    376$\pm$& 224 &  307$\pm$&       50 &    308$\pm$ &   45 &$\star$/?\\
17&L37& 43855$\pm$& 162 &  17383$\pm$& 168 &  \nodata &          &  31100$\pm$ &   82 &$\ominus$/??\\
18&L26& 22836$\pm$& 339 &  22952$\pm$& 333 &  \nodata &          &  22895$\pm$ &  168 &$\ominus$\\
19&L39& 27415$\pm$& 366 &  22664$\pm$& 251 &  \nodata &          &  24184$\pm$ &  146 &$\ominus$/??\\
20&L52&    87$\pm$& 287 &    168$\pm$& 217 &  \nodata &          &    139$\pm$ &  122 &$\star$/??\\
21&L17&131653$\pm$& 187 & 131659$\pm$&  79 &  \nodata &          & 131658$\pm$ &   51 &$\ominus$\\
22&L38& 24447$\pm$& 135 &  26512$\pm$& 335 &  \nodata &          &  24735$\pm$ &   89 &$\ominus$/?\\
\enddata

\tablenotetext{a}{Ordinal number of slits in each mask. Along the
  alignment of both masks slit 1 is in the most north-western part of
  the field-of-view as shown in Figure \ref{ps:fov}.}
\tablenotetext{b}{Object identification marker. The code corresponds
  to markers of spectroscopically identified globular clusters in
  Figure \ref{ps:fov}.} \tablenotetext{c}{Classification category of
  the spectrum. It is: $\star$ -- star, $\ominus$ -- galaxy, and
  $\oplus$ -- globular cluster. ``?'' indicates that the spectrum
  shows only slight features of a typical spectrum in the
  classification category while ``??'' indicates that due to too low
  S/N no classification is possible by visual inspection.}
\tablecomments{Summary of all radial velocities including ``internal''
  cross-correlation errors and statistical errors from line shift
  measurements. All radial velocities are given in km s$^{-1}$.
  $V_{\rm rad}^{\rm lines}$ is a mean radial velocity calculated from
  redshifts of the element lines: \ion{Ca}{2} K (3933\AA ),
  \ion{Ca}{2} H (3967\AA ), H$\delta$ (4101\AA ), \ion{Ca}{1} (4226\AA
  ), H$\gamma$ (4340\AA ), \ion{Fe}{1} (4383\AA ), \ion{Fe}{1}
  (4528\AA ), H$\beta$ (4861\AA ), \ion{Mg}{1} (5183\AA ), \ion{Fe}{1}
  (5270\AA ) \citep[][rest wavelengths in
  parentheses]{apq73}.}
\end{deluxetable}
\clearpage
\begin{deluxetable}{lcccc}
\tabletypesize{\scriptsize}
\tablecaption{Lower Mass Limits of NGC 3115 DW1 \label{tab:massestimate}}
\tablewidth{0pt}
\tablehead{
&\multicolumn{4}{c}{masses in $10^{10}$M$_\odot$}\\\cline{2-5}\\
\colhead{estimator} & \colhead{$M_{\rm gal}$} & 
\colhead{$\sigma_{\rm intr}$\tablenotemark{a}} & 
\colhead{$\sigma_{\rm stat}$\tablenotemark{b}} & 
\colhead{$\sigma_{\rm dist}$\tablenotemark{c}}  
}
\startdata
VME             & 2.11 &\nodata&  0.04 &$^{+0.95}_{-0.45}$\\
PME$_{\rm i}$   & 4.77 & 2.28  &  0.16 &$^{+2.17}_{-1.00}$\\
PME$_{\rm r}$   & 9.54 & 5.05  &  0.32 &$^{+4.26}_{-2.00}$\\
PME$_{\rm m}$   & 7.15 & 2.77  &  0.24 &$^{+3.25}_{-1.49}$\\
PME$_{\rm t}$   & 3.18 & 1.41  &  0.11 &$^{+1.42}_{-0.67}$\\
\enddata


\tablenotetext{a}{Individual intrinsic uncertainty of the
  mass-estimation code. Note that for VME there is no analytical
  variance formula available.}
\tablenotetext{b}{Standard uncertainty of the mass estimate due to the
  limited sample size as numerically determined in Bootstrap tests.} 
\tablenotetext{c}{Systematic uncertainty due to potential distance
  error.}
\tablecomments{Lower mass estimate for NGC 3115 DW1 out to a radius of
  $r\leq189.4$\arcsec\ or $R\leq10.1$ kpc. All masses are given in
  units of $10^{10} M_\odot$. Different orbit characteristics of the
  globular clusters were adopted for the mass estimate using the
  Projected Mass Estimator (PME): PME$_{\rm i}$ adopts isotropic
  orbits, PME$_{\rm r}$ adopts radial orbits, PME$_{\rm m}$ assumes
  mixed while PME$_{\rm t}$ adopts tangential globular-cluster orbits
  \citep[for details see][]{bahcall81}.}
\end{deluxetable}
\begin{deluxetable}{llcccc}
\tabletypesize{\scriptsize}
\tablecaption{Radial Mass Distribution \label{tab:massrad}}
\tablewidth{0pt}
\tablehead{
\multicolumn{2}{c}{} & \multicolumn{3}{c}{masses in
  $10^{10}$M$_\odot$}&\\\cline{3-5}\\
\colhead{$r_i$} & \colhead{$R_i$} &
\colhead{$M_i$\tablenotemark{a}} &
\colhead{$\sigma_{{\rm intr},i}$\tablenotemark{b}} &
\colhead{$\sigma_{{\rm stat},i}$\tablenotemark{c}} &
\colhead{\# of GCs\tablenotemark{d}} 
}
\startdata
$r <189.4$\arcsec\ & $R < 10.1$ kpc& 4.77 & 2.28 & 0.16 & 7\\
$r <161.4$\arcsec\ & $R <  8.6$ kpc& 4.55 & 2.35 & 0.25 & 6\\
$r < 56.3$\arcsec\ & $R <  3.0$ kpc& 1.82 & 1.03 & 0.08 & 5\\
$r < 41.8$\arcsec\ & $R <  2.2$ kpc& 1.61 & 1.02 & 0.06 & 4\\
\enddata


\tablenotetext{a}{Mass estimate for the $i$-th projected radial
  distance evaluated with PME for isotropic globular-cluster
  orbits. The masses are given in units of $10^{10}\:M_\odot$.}
\tablenotetext{b}{Internal uncertainty of the mass-estimation code.}
\tablenotetext{c}{Standard uncertainty from Bootstrap tests.}
\tablenotetext{d}{Number of globular clusters included in the data
  sample which was used to estimate the mass at the $i$-th radius.}
\end{deluxetable}
\clearpage
\begin{deluxetable}{lrrcrrc}
\tabletypesize{\scriptsize}
\tablecaption{Abundances of individual globular clusters and the
  entire system \label{tab:abundmag}}
\tablewidth{0pt}
\tablehead{
\colhead{}   &
\multicolumn{3}{c}{D46} &
\multicolumn{3}{c}{D14} \\
\cline{2-4} \cline{5-7}\\
\colhead{line/band} &
\colhead{$I$ [mag]} & \colhead{[Fe/H]$_i$} & \colhead{$w_i$} &
\colhead{$I$ [mag]} & \colhead{[Fe/H]$_i$} & \colhead{$w_i$} 
}
\startdata
CNB     &$ 0.166\pm0.781$ &$-0.847$& 0.195 &$ 0.242\pm 0.773$ &$-0.349$& 0.197\\
H+K     &$-0.152\pm0.814$ &$-4.430$& 0.155 &$ 0.459\pm 0.521$ &$ 0.418$& 0.242\\
CNR     &$ 0.013\pm0.414$ &$-1.098$& 0.328 &$ 0.091\pm 0.287$ &$-0.522$& 0.474\\
CH=Gband&$-0.121\pm0.699$ &$-3.842$& 0.126 &$ 0.229\pm 0.358$ &$ 0.154$& 0.246\\
H$\beta$&$ 0.569\pm0.389$ &$-0.431$&\nodata&$-0.007\pm 0.215$ &$-1.007$&\nodata\\ 
MgH     &$ 0.074\pm0.191$ &$-0.309$& 0.257 &$ 0.120\pm 0.132$ &$ 0.620$& 0.372\\
MgG     &$ 0.352\pm0.287$ &$-0.648$&\nodata&$ 0.041\pm 0.181$ &$-0.959$&\nodata\\
Mg2     &$ 0.001\pm0.204$ &$-2.198$& 0.496 &$ 0.084\pm 0.145$ &$-1.377$& 0.698\\
Mgb     &$ 0.185\pm0.380$ &$-0.815$&\nodata&$ 0.004\pm 0.217$ &$-0.996$&\nodata\\
Fe52    &$-0.131\pm0.336$ &$-4.755$& 0.146 &$ 0.102\pm 0.222$ &$-0.010$& 0.221\\
\tableline
$\langle$[Fe/H]$\rangle$ &
\multicolumn{3}{c}{$-0.84\pm0.82$ dex}&
\multicolumn{3}{c}{$-0.86\pm0.39$ dex}\\ 
\multicolumn{7}{c}{}\\

&\multicolumn{3}{c}{D15} &
\multicolumn{3}{c}{D7} \\
\cline{2-4} \cline{5-7}\\
CNB      &$ 0.132\pm 0.245$ &$-1.072$& 0.619 &$ 0.093\pm 0.204$ &$-1.327$& 0.745\\
H+K      &$ 0.308\pm 0.203$ &$-0.782$& 0.620 &$ 0.400\pm 0.162$ &$-0.050$& 0.777\\
CNR      &$ 0.111\pm 0.151$ &$-0.378$& 0.898 &$ 0.070\pm 0.110$ &$-0.675$& 1.242\\
CH=Gband &$ 0.103\pm 0.178$ &$-1.281$& 0.494 &$ 0.187\pm 0.116$ &$-0.316$& 0.756\\
H$\beta$ &$-0.010\pm 0.113$ &$-1.010$&\nodata&$ 0.074\pm 0.079$ &$-0.926$&\nodata\\
MgH      &$ 0.023\pm 0.068$ &$-1.371$& 0.726 &$ 0.086\pm 0.047$ &$-0.065$& 1.050\\
MgG      &$-0.040\pm 0.084$ &$-1.040$&\nodata&$-0.007\pm 0.054$ &$-1.007$&\nodata\\
Mg2      &$ 0.069\pm 0.078$ &$-1.528$& 1.295 &$ 0.098\pm 0.056$ &$-1.235$& 1.796\\
Mgb      &$ 0.010\pm 0.105$ &$-0.990$&\nodata&$ 0.060\pm 0.066$ &$-0.940$&\nodata\\
Fe52     &$ 0.080\pm 0.105$ &$-0.461$& 0.467 &$ 0.090\pm 0.070$ &$-0.245$& 0.701\\
\tableline
$\langle$[Fe/H]$\rangle$ &
\multicolumn{3}{c}{$-1.01\pm0.20$ dex}&
\multicolumn{3}{c}{$-0.91\pm0.23$ dex}\\ 
\multicolumn{7}{c}{}\\

&\multicolumn{3}{c}{D25} & 
\multicolumn{3}{c}{L1} \\
\cline{2-4} \cline{5-7}\\
\colhead{line/band} &
\colhead{$I$ [mag]} & \colhead{[Fe/H]$_i$} & \colhead{$w_i$} &
\colhead{$I$ [mag]} & \colhead{[Fe/H]$_i$} & \colhead{$w_i$} \\
\tableline
CNB      &$ 0.026\pm 0.363$ &$-1.770$& 0.419 &$ 0.286\pm 0.178$ &$-0.059$& 0.854\\ 
H+K      &$ 0.163\pm 0.316$ &$-1.933$& 0.399 &$ 0.279\pm 0.159$ &$-1.016$& 0.791\\ 
CNR      &$-0.031\pm 0.206$ &$-1.422$& 0.660 &$ 0.083\pm 0.103$ &$-0.585$& 1.323\\ 
CH=Gband &$ 0.105\pm 0.243$ &$-1.259$& 0.363 &$ 0.057\pm 0.108$ &$-1.805$& 0.818\\ 
H$\beta$ &$ 0.022\pm 0.185$ &$-0.978$&\nodata&$ 0.023\pm 0.063$ &$-0.977$&\nodata\\ 
MgH      &$-0.002\pm 0.097$ &$-1.891$& 0.506 &$ 0.055\pm 0.039$ &$-0.701$& 1.249\\
MgG      &$-0.001\pm 0.121$ &$-1.001$&\nodata&$-0.013\pm 0.042$ &$-1.013$&\nodata\\
Mg2      &$-0.024\pm 0.107$ &$-2.449$& 0.944 &$ 0.097\pm 0.049$ &$-1.246$& 2.074\\
Mgb      &$-0.041\pm 0.155$ &$-1.041$&\nodata&$-0.015\pm 0.053$ &$-1.015$&\nodata\\
Fe52     &$ 0.042\pm 0.159$ &$-1.238$& 0.308 &$ 0.082\pm 0.058$ &$-0.420$& 0.847\\
\tableline
$\langle$[Fe/H]$\rangle$ &
\multicolumn{3}{c}{$-1.11\pm0.37$ dex}&
\multicolumn{3}{c}{$-0.98\pm0.19$ dex}\\
\multicolumn{7}{c}{}\\

&\multicolumn{3}{c}{L63} & 
\multicolumn{3}{c}{$\sum$}\\
\cline{2-4} \cline{5-7} \\
\colhead{line/band} &
\colhead{$I$ [mag]} & \colhead{[Fe/H]$_i$} & \colhead{$w_i$} &
\colhead{$I$ [mag]} & \colhead{[Fe/H]$_i$} & \colhead{$w_i$} \\
\tableline
CNB      &$-0.603\pm 3.794$ &$-5.903$& 0.040 & $0.123\pm0.125$&$-1.133$& 1.215\\
H+K      &$ 0.022\pm 0.790$ &$-3.048$& 0.159 & $0.274\pm0.110$&$-1.052$& 1.147\\
CNR      &$ 0.237\pm 0.499$ &$ 0.552$& 0.273 & $0.072\pm0.085$&$-0.664$& 1.608\\
CH=Gband &$ 0.596\pm 0.634$ &$ 4.347$& 0.139 & $0.128\pm0.080$&$-0.999$& 1.104\\
H$\beta$ &$ 0.357\pm 0.445$ &$-0.643$&\nodata& $0.067\pm0.055$&$-0.933$&\nodata\\
MgH      &$ 0.011\pm 0.220$ &$-1.620$& 0.223 & $0.056\pm0.035$&$-0.681$& 1.401\\
MgG      &$-0.033\pm 0.286$ &$-1.033$&\nodata& $0.008\pm0.035$&$-0.992$&\nodata\\
Mg2      &$ 0.097\pm 0.251$ &$-1.248$& 0.402 & $0.073\pm0.044$&$-1.486$& 2.275\\
Mgb      &$ 0.235\pm 0.337$ &$-0.765$&\nodata& $0.029\pm0.043$&$-0.971$&\nodata\\
Fe52     &$ 0.040\pm 0.336$ &$-1.277$& 0.146 & $0.067\pm0.051$&$-0.727$& 0.955\\
\tableline
$\langle$[Fe/H]$\rangle$ &
\multicolumn{3}{c}{$-0.81\pm0.81$ dex}&
\multicolumn{3}{c}{$-0.97\pm0.11$ dex}\\
\enddata

\tablecomments{Abundances measured with the Lick/IDS passbands defined
  by \citet{brodie90}. The line strengths are given in mag. With the
  metallicity calibrations of \citeauthor{brodie90} we derive for the
  $i$-th absorption line an individual metallicity [Fe/H]$_i$.
  According to the statistical significance of each individual
  metallicity indicator its weighting $w_i$ is given. The weighted
  mean metallicity of each globular cluster and of the entire GCS
  (indicated as $\sum$) is given at the bottom line. Lines which were
  found to be very poor or redundant metallicity indicators
  \citep{brodie90} have no weighting assigned and are excluded from
  the averaging process. The errors include both Poisson flux
  uncertainties and metallicity-calibration errors.}
\end{deluxetable}
\clearpage
\begin{deluxetable}{lcc}
\tabletypesize{\scriptsize}
\tablecaption{Comparison of [Fe/H] from photometry and spectroscopy
  \label{tab:metalcomp}}
\tablewidth{0pt}
\tablehead{
\colhead{cluster} & \colhead{[Fe/H]$_{\rm spec}$} &
\colhead{[Fe/H]$_{\rm phot}$}
}
\startdata
D7  & $-0.91\pm$0.23 & $-0.66\pm$0.20 \\
D14 & $-0.86\pm$0.39 & $-0.61\pm$0.20 \\
D15 & $-1.01\pm$0.20 & $-1.11\pm$0.20 \\
D25 & $-1.11\pm$0.37 & $-1.66\pm$0.30 \\
D46 & $-0.84\pm$0.82 & $-1.41\pm$0.45 \\
L1  & $-0.98\pm$0.19 &  \nodata \\
L63 & $-0.81\pm$0.81 &  \nodata \\\tableline
\multicolumn{3}{c}{}\\
&$\langle$[Fe/H]$\rangle_{\rm spec}=-0.97\pm0.11$
&$\langle$[Fe/H]$\rangle_{\rm phot}=-0.93\pm0.11$\\
\multicolumn{3}{c}{}\\
\enddata

\tablecomments{The spectroscopical mean metallicity was obtained from
  the mean spectrum of 7 globular clusters. Exactly the same values
  can be calculated by weighted averaging of individual globular
  cluster metallicities. The photometric mean metallicity is a
  weighted mean of the above values. The error of
  $\langle$[Fe/H]$\rangle_{\rm phot}$ includes only the photometric
  uncertainty of the $(B-V)$ color. It does not include the error of
  the color-metallicity calibration (c.f. equation
  \ref{eq:colormetallin}).}
\end{deluxetable}
\clearpage
\begin{deluxetable}{rcccccrrrp{0.005cm}r}
\tabletypesize{\scriptsize}
\tablecaption{Spectroscopic and photometric data of all
  spectroscopically studied objects \label{tab:specphot}} 
\tablewidth{0pt}
\tablehead{
\colhead{Label} & 
\colhead{RA\tablenotemark{a}} & 
\colhead{Dec\tablenotemark{a}} & 
\colhead{$V$\tablenotemark{b}} &
\colhead{$(V-I)$\tablenotemark{c}}  & 
\colhead{$(B-V)$\tablenotemark{d}} & 
\colhead{$r$\tablenotemark{e}} & 
\colhead{$R$\tablenotemark{e}} & 
\multicolumn{3}{c}{$\langle V_{\rm rad}\rangle$\tablenotemark{f}}\\

\colhead{} & 
\colhead{[deg] (J2000)} & 
\colhead{[deg] (J2000)} & 
\colhead{[mag]} &
\colhead{[mag]}  & 
\colhead{[mag]} & 
\colhead{[\arcsec ]} & 
\colhead{[pc]} & 
\multicolumn{3}{c}{[km s$^{-1}$]}

}
\startdata
\multicolumn{11}{c}{}\\
\multicolumn{11}{c}{mask A}\\\tableline
\multicolumn{11}{c}{}\\
L15 & 151.46547275 &$-$8.02573829 &      \nodata     &     \nodata     &     \nodata     & 215.2 & 11474 & 70407&$\pm$&106 \\
 L3 & 151.44513389 &$-$8.03275119 &      \nodata     &     \nodata     &     \nodata     & 197.4 & 10525 &    69&$\pm$& 57 \\
L27 & 151.45595107 &$-$8.01588459 &      \nodata     &     \nodata     &     \nodata     & 166.1 &  8856 & 25235&$\pm$&105 \\
L11 & 151.47453033 &$-$8.00005692 &      \nodata     &     \nodata     &     \nodata     & 190.1 & 10139 &    84&$\pm$& 58 \\
L18 & 151.45373013 &$-$8.00788268 & 21.65 $\pm$ 0.05 &     \nodata     & 0.74 $\pm$ 0.06 & 140.2 &  7475 &   199&$\pm$& 42 \\
 L7 & 151.43413151 &$-$8.01463921 &      \nodata     &     \nodata     &     \nodata     & 122.8 &  6549 & $-$60&$\pm$& 81 \\
L16 & 151.43857103 &$-$8.00897480 &      \nodata     &     \nodata     &     \nodata     & 109.7 &  5850 &$-$196&$\pm$& 85 \\
 D8 & 151.44176680 &$-$8.00230439 & 21.70 $\pm$ 0.02 &     \nodata     & 1.47 $\pm$ 0.06 &  96.0 &  5117 &    86&$\pm$& 59 \\
D21 & 151.42597410 &$-$8.00187100 & 21.64 $\pm$ 0.03 & 1.37 $\pm$ 0.06 & 0.70 $\pm$ 0.04 &  72.1 &  3845 &\multicolumn{3}{c}{\nodata} \\
D10 & 151.45237834 &$-$7.98267887 & 20.97 $\pm$ 0.02 &     \nodata     & 1.10 $\pm$ 0.03 &  99.6 &  5312 &   110&$\pm$& 39 \\
D58 & 151.43598000 &$-$7.99663820 & 23.22 $\pm$ 0.11 & 2.09 $\pm$ 0.10 & 1.09 $\pm$ 0.18 &  67.1 &  3579 &  8766&$\pm$& 89 \\
D73 & 151.43584940 &$-$7.98355023 & 23.38 $\pm$ 0.09 &     \nodata     & 0.64 $\pm$ 0.15 &  41.1 &  2190 &  5015&$\pm$&128 \\
D20 & 151.42182290 &$-$7.98467580 & 21.72 $\pm$ 0.04 & 1.55 $\pm$ 0.15 & 0.65 $\pm$ 0.06 &  13.7 &   731 & 22562&$\pm$&103 \\
D24 & 151.43774990 &$-$7.97262110 & 21.78 $\pm$ 0.04 & 1.28 $\pm$ 0.07 & 0.67 $\pm$ 0.07 &  58.0 &  3093 &   157&$\pm$&112 \\
D14 & 151.42099595 &$-$7.97540485 & 21.34 $\pm$ 0.03 &     \nodata     & 0.90 $\pm$ 0.04 &  26.4 &  1409 &   468&$\pm$& 71 \\
 D6 & 151.43119970 &$-$7.96549650 & 20.28 $\pm$ 0.02 & 1.58 $\pm$ 0.03 & 0.50 $\pm$ 0.03 &  63.8 &  3401 &    24&$\pm$& 32 \\
 D1 & 151.40556550 &$-$7.97613822 & 18.85 $\pm$ 0.01 &     \nodata     & 0.73 $\pm$ 0.01 &  70.5 &  3757 &   147&$\pm$& 39 \\
 D5 & 151.41731295 &$-$7.96373779 & 20.32 $\pm$ 0.01 &     \nodata     & 0.60 $\pm$ 0.02 &  70.2 &  3742 &   131&$\pm$& 36 \\
L37 & 151.42092628 &$-$7.95714247 &      \nodata     &     \nodata     &     \nodata     &  90.0 &  4800 & 24560&$\pm$&102 \\
L44 & 151.43991405 &$-$7.94358566 &      \nodata     &     \nodata     &     \nodata     & 148.5 &  7922 & 22998&$\pm$&108 \\
L14 & 151.43728995 &$-$7.93698967 &      \nodata     &     \nodata     &     \nodata     & 168.0 &  8961 &    52&$\pm$& 53 \\
L51 & 151.42385948 &$-$7.93980723 &      \nodata     &     \nodata     &     \nodata     & 151.5 &  8082 &    47&$\pm$& 96 \\
L1  & 151.41836728 &$-$7.93746430 &      \nodata     &     \nodata     &     \nodata     & 161.4 &  8609 &   420&$\pm$& 29 \\
L63 & 151.40348832 &$-$7.93354949 &      \nodata     &     \nodata     &     \nodata     & 189.4 & 10102 &   605&$\pm$& 74 \\
\multicolumn{11}{c}{}\\
\multicolumn{11}{c}{mask B}\\\tableline
\multicolumn{11}{c}{}\\
 L8 & 151.42908177 &$-$8.03330443 &      \nodata     &     \nodata     &     \nodata     & 185.8 &  9908 &  99444&$\pm$& 65 \\
 L4 & 151.45874258 &$-$8.02204502 &      \nodata     &     \nodata     &     \nodata     & 189.3 & 10095 &     49&$\pm$& 60 \\
 L6 & 151.43119125 &$-$8.02568813 &      \nodata     &     \nodata     &     \nodata     & 159.5 &  8504 &\multicolumn{3}{c}{\nodata}\\
L23 & 151.44361588 &$-$8.01586187 &      \nodata     &     \nodata     &     \nodata     & 140.1 &  7469 & $-$293&$\pm$&106 \\
 D9 & 151.43313299 &$-$8.00698070 & 21.63 $\pm$ 0.02 &     \nodata     & 1.52 $\pm$ 0.06 &  95.5 &  5090 &     24&$\pm$& 61 \\
D36 & 151.44962050 &$-$7.99643065 & 22.86 $\pm$ 0.07 &     \nodata     & 1.70 $\pm$ 0.17 & 103.9 &  5539 &   $-$4&$\pm$& 73 \\
D26 & 151.44384880 &$-$7.98342710 & 22.49 $\pm$ 0.07 & 1.25 $\pm$ 0.11 & 1.43 $\pm$ 0.13 &  69.4 &  3700 &$-$2567&$\pm$&109 \\
 D2 & 151.43874410 &$-$7.99131620 & 20.32 $\pm$ 0.02 & 2.69 $\pm$ 0.04 & 1.60 $\pm$ 0.03 &  61.2 &  3264 &  $-$11&$\pm$& 59 \\
D25 & 151.43899620 &$-$7.98802330 & 21.78 $\pm$ 0.04 & 1.27 $\pm$ 0.07 & 0.69 $\pm$ 0.06 &  56.3 &  3005 &    532&$\pm$& 60 \\
D46 & 151.42821583 &$-$7.98176333 & 22.53 $\pm$ 0.05 &     \nodata     & 0.74 $\pm$ 0.09 &  13.4 &   716 &    623&$\pm$&106 \\
 D3 & 151.44378760 &$-$7.97441980 & 20.74 $\pm$ 0.02 & 3.32 $\pm$ 0.03 & 1.53 $\pm$ 0.04 &  74.0 &  3947 &   $-$5&$\pm$& 56 \\
D15 & 151.43225130 &$-$7.97337860 & 21.30 $\pm$ 0.02 & 1.52 $\pm$ 0.05 & 0.80 $\pm$ 0.04 &  41.4 &  2208 &    681&$\pm$& 30 \\
 D7 & 151.41813758 &$-$7.97210601 & 20.53 $\pm$ 0.03 &     \nodata     & 0.89 $\pm$ 0.04 &  41.8 &  2231 &    703&$\pm$& 25 \\
D42 & 151.44129540 &$-$7.96256740 & 22.79 $\pm$ 0.07 & 1.56 $\pm$ 0.10 & 1.63 $\pm$ 0.18 &  91.9 &  4902 &  58100&$\pm$&123 \\
L36 & 151.45635892 &$-$7.95500702 &      \nodata     &     \nodata     &     \nodata     & 149.4 &  7966 &  26386&$\pm$& 96 \\
D41 & 151.40979676 &$-$7.96307913 & 22.19 $\pm$ 0.04 &     \nodata     & 0.49 $\pm$ 0.06 &  85.6 &  4563 &    308&$\pm$& 45 \\
L37 & 151.42092628 &$-$7.95714247 &      \nodata     &     \nodata     &     \nodata     &  90.0 &  4800 &  31100&$\pm$& 82 \\
L26 & 151.44282349 &$-$7.94889087 &      \nodata     &     \nodata     &     \nodata     & 135.7 &  7236 &  22895&$\pm$&168 \\
L39 & 151.42936599 &$-$7.94366459 &      \nodata     &     \nodata     &     \nodata     & 138.8 &  7400 &  24184&$\pm$&146 \\
L52 & 151.41051686 &$-$7.94476283 &      \nodata     &     \nodata     &     \nodata     & 142.6 &  7606 &    139&$\pm$&122 \\
L17 & 151.45223278 &$-$7.92867126 &      \nodata     &     \nodata     &     \nodata     & 215.7 & 11504 & 131658&$\pm$& 51 \\
L38 & 151.42761022 &$-$7.93031858 &      \nodata     &     \nodata     &     \nodata     & 186.0 &  9921 &  24735&$\pm$& 89 \\
\enddata

\tablenotetext{a}{Both RA and Dec have an accuracy of $\delta$(RA,
  Dec)$\leq2$\arcsec . The center of NGC 3115 DW1 has the coordinates
  RA$=151.4232917$ and Dec$=7.981527778$.}
\tablenotetext{b}{All $V$ magnitudes were taken from
  \citet{durrell96}.}

\tablenotetext{c}{$V$ in $(V-I)$ was obtained from \citet{durrell96}
  while the $I$ magnitude was taken from our HST photometry. Note that
  the $(V-I)$ color suffers large uncertainty due to low S/N of the
  F814W images. The $(V-I)$ color is not corrected for galactic
  extinction and has to be de-reddened (E$_{(B-V)}=0.052$ mag
  \citep{schlegel98}; E$_{(V-I)}=1.3\cdot$E$_{(B-V)}$ \citep{dean78}
  yields E$_{(V-I)}=0.068$ mag).}
\tablenotetext{d}{$(B-V)$ was obtained from \citet{durrell96}.}
\tablenotetext{e}{Projected radial distance to the center of NGC 3115
  DW1.}
\tablenotetext{f}{Mean radial velocity calculated from
  cross-correlations and line shifts.}
\end{deluxetable}

\clearpage

\end{document}